\documentclass[twocolumn]{aastex631}
\pdfoutput=1
\usepackage{graphicx, url}
\usepackage{enumerate, enumitem}
\usepackage{amssymb, amsmath, amsfonts, upgreek}
\usepackage{gensymb}
\usepackage{mathtools}
\usepackage{xspace}
\usepackage{xcolor}
\usepackage[normalem]{ulem}
\usepackage{amsmath}
\usepackage{amssymb}
\usepackage{natbib}
\usepackage{graphicx}
\usepackage{xcolor}
\usepackage{hyperref}
 
\newcommand{\Msun}{M_\odot}
\newcommand{\ameas}{\alpha_{\rm meas}}
\newcommand{\amax}{\alpha_{\rm max}}
\newcommand{\aratio}{\mathcal{R}}
\newcommand{\masyr}{{\rm mas\,yr}^{-2}}
\newcommand{\gaia}{\textit{Gaia}}
\newcommand{\Hipparcos}{\textit{Hipparcos}}
\newcommand{\orvara}{\texttt{orvara}}

\newcommand{\nnss}{338,215\xspace}        % NSS acceleration solutions with a parallax
\newcommand{\nacca}{246,947\xspace}       % ... Acceleration7
\newcommand{\naccb}{91,268\xspace}        % ... Acceleration9
\newcommand{\nwdscat}{157,391\xspace}     % WDS entries
\newcommand{\nmatch}{2,468\xspace}        % positional matches < 2"
\newcommand{\nuniq}{2,466\xspace}         % de-duplicated matches (the full sample)
\newcommand{\nmagpair}{2,455\xspace}      % both WDS magnitudes finite
\newcommand{\pmagpair}{99.6\xspace}       % ... per cent
\newcommand{\nmissmtwo}{11\xspace}        % no WDS m2, valid primary
\newcommand{\nflame}{1,453\xspace}
\newcommand{\pflame}{58.9\xspace}
\newcommand{\nhipmass}{111\xspace}
\newcommand{\phipmass}{4.5\xspace}
\newcommand{\ntwomassin}{902\xspace}
\newcommand{\ntwomass}{896\xspace}
\newcommand{\ptwomass}{36.3\xspace}
\newcommand{\nmone}{2,460\xspace}
\newcommand{\pmone}{99.8\xspace}
\newcommand{\nnomone}{6\xspace}
\newcommand{\pnomone}{0.2\xspace}
\newcommand{\ndmwds}{2,455\xspace}
\newcommand{\pdmwds}{99.6\xspace}

\newcommand{\ndmnone}{11\xspace}
\newcommand{\pdmnone}{0.4\xspace}
\newcommand{\nmtworoot}{2,449\xspace}
\newcommand{\pmtworoot}{99.3\xspace}
\newcommand{\nswapped}{Forty-eight\xspace}  % Delta m < 0 in the full sample
\newcommand{\nequal}{55\xspace}           % equal-magnitude pairs, alpha_max = 0
\newcommand{\nsepzero}{28\xspace}         % sep1 = 0
\newcommand{\nnoratio}{84\xspace}         % = nuniq - nfinite
\newcommand{\nfinite}{2,382\xspace}       % finite ratio
\newcommand{\nhigher}{2,336\xspace}       % R > 10 among the finite
\newcommand{\phigher}{98.1\xspace}
\newcommand{\medRfull}{3.7\xspace}        % median R of the finite sample, x 10^5
\newcommand{\fpcfull}{0.50\xspace}
\newcommand{\fpcplaus}{0.72\xspace}
\newcommand{\nuncorr}{65\xspace}          % plausible without the photocenter correction
\newcommand{\nplaus}{45\xspace}           % the plausible sample
\newcommand{\nplausa}{30\xspace}
\newcommand{\nplausb}{15\xspace}   % Acc7 / Acc9
\newcommand{\nplauship}{22\xspace}
\newcommand{\nplausnohip}{23\xspace}
\newcommand{\Nplauship}{Twenty-two\xspace}
\newcommand{\medR}{2.96\xspace}
\newcommand{\meanR}{3.28\xspace}
\newcommand{\minR}{0.12\xspace}
\newcommand{\maxR}{8.00\xspace}
\newcommand{\nexplain}{41\xspace}
\newcommand{\nrhoquant}{26\xspace}        % rho_1 = 0.1" among the plausible
\newcommand{\fRbelowone}{22\xspace}
\newcommand{\fRbelowmed}{9\xspace}
\newcommand{\medRiso}{4.6\xspace}
\newcommand{\medsepAU}{8.7\xspace}
\newcommand{\minsepAU}{3.4\xspace}
\newcommand{\maxsepAU}{20.4\xspace}
\newcommand{\meddpc}{56\xspace}
\newcommand{\nredflag}{18\xspace}
\newcommand{\nkflag}{12\xspace}
\newcommand{\nbandflag}{30\xspace}
\newcommand{\pearsona}{0.98\xspace}
\newcommand{\pearsonb}{0.22\xspace}
\newcommand{\nfit}{17\xspace}
\newcommand{\Nfit}{seventeen\xspace}
\newcommand{\nfitplaus}{16\xspace}        % fits among the 45 (16206+4535 is outside)

\newcommand{\nsparse}{23\xspace}

\newcommand{\Rboundary}{12.3\xspace}
\newcommand{\Rboundaryunc}{9.99\xspace}
\newcommand{\fpc}{f_{\rm pc}\xspace}
\newcommand{\ntest}{43\xspace}
\newcommand{\nuntest}{two\xspace}
\newcommand{\nalign}{33\xspace}
\newcommand{\palign}{77\xspace}
\newcommand{\medoff}{15\fdg0\xspace}
\newcommand{\binp}{3.0\times10^{-4}\xspace}
\newcommand{\ksD}{0.355\xspace}
\newcommand{\ksp}{2.4\times10^{-5}\xspace}
\newcommand{\nctrlR}{2,336\xspace}
\newcommand{\nctrl}{2,331\xspace}
\newcommand{\nctrlalign}{1,166\xspace}
\newcommand{\pctrl}{50.0\xspace}
\newcommand{\medoffctrl}{45\fdg0\xspace}
\newcommand{\ksDctrl}{0.013\xspace}
\newcommand{\kspctrl}{0.80\xspace}
\newcommand{\nfittest}{16\xspace}
\newcommand{\nfitalign}{12\xspace}
\newcommand{\nnofittest}{27\xspace}
\newcommand{\nnofitalign}{21\xspace}
\newcommand{\nsparsetest}{21\xspace}
\newcommand{\nsparsealign}{17\xspace}

\newcommand{\nbinned}{2,356\xspace}
\newcommand{\nunbinned}{19\xspace}

\begin{document}
 
\title{Validating Gaia DR3 Astrometric Accelerations with Close Visual Binaries}
 
\author[0000-0003-0115-547X]{Qier An}
\affiliation{Department of Physics and Astronomy, Johns Hopkins University, Baltimore, MD 21218, USA}
 
\author[0000-0003-2630-8073]{Timothy D. Brandt}
\affiliation{Space Telescope Science Institute, Baltimore, MD 21218, USA}
 
\correspondingauthor{Qier An}
\email{qan4@jh.edu}  
 
\begin{abstract}
We validate a subset of the astrometric accelerations reported in \gaia{} DR3 non-single-star solutions by their consistency with the gravitational pull of known, resolved companions. Cross-matching \nnss DR3 acceleration solutions against the Washington Double Star (WDS) catalog of visual binaries yields \nuniq unique pairs within $2''$. For each pair we compare the measured acceleration with $\alpha_{\rm max}$, the Newtonian bound on the acceleration the companion can impart on the \gaia{}-tracked photocenter. For \phigher{}\% of the pairs the measured acceleration exceeds this bound by more than a factor of 10, so the cataloged companion cannot be responsible and the perturber must be a closer, unresolved one. The remaining \nplaus plausible wide binaries are close pairs at the WDS resolution limit. We fit orbits with an updated \orvara{}, extended to accept the DR3 acceleration and jerk coefficients as direct constraints and released with \texttt{htof} as part of this work. The fits converge for 16 of the \nplaus and one boundary case, the rest having too few resolved measures or a perturber other than the cataloged pair, and 13 of the 17 carry enough data to test the coefficients. Eleven of the 13 statistically accommodate the reported accelerations, for seven- and nine-parameter solutions alike, and the dynamical companion masses agree with the photometric estimates where the photocenter conversion is well determined. Independently of the fits, \nalign of the \ntest plausible systems with a usable position angle (\palign{}\%) have the acceleration directed within $45^\circ$ of the primary--companion axis, against 50\% for a control sample: DR3's measured accelerations point in the directions of the cataloged companions.

% Abstract bookkeeping (checked against wds_gaia_full.csv, direction_check_*.csv and Table 3, 2026-11-20):
% 98.1% is 2,336 of the 2,382 pairs with a finite ratio, not of 2,466: 84 pairs have no ratio (6 no M1, 28 no sep1, 55 equal-magnitude, 5 overlapping).
% 2,382 = 2,336 (R > 10) + 45 (plausible) + 1 below the window (18267-3024, R = 0.08; Mode B fit attempted, rejected, Section 3.4; not among the 17 fits).
% The 45 plausible: 16 converged fits + 6 excluded (4 with a perturber other than the cataloged pair, 2 with an acceleration not constant across the baseline) + 23 too sparse WDS data.
% The 16 fits plus the boundary case 16206+4535 (R = 12.3, just outside the window, Table 2 note a) are 17 fits: 9 Mode B + 8 Mode A.
% 13 of the 17 test the DR3 coefficients (9 Mode B + 4 Mode A Acc9); the other 4 Mode A Acc7 fits have two epochs, 6 data for 7 parameters, and test nothing.
% 11 of the 13 reproduce the coefficients, i.e. the fit matches every data set with no systematic residual (chi2_G/nu <= 0.58 for ten of them, 3.82 for 23464-2302).
% The two failures: 14382+1402 (chi2_G/nu = 16.4, acceleration reproduced but not the jerk, Table 2 note d) and 19169-4348 (chi2_G/nu = 1.64 with chi2_rho/N = 9.0, P = 3.3 yr ~ the DR3 baseline).
% 45 -> 43 in the direction test: 08423+1322 and 20335-1333 have only occultation timings, no position angle. 33 of 43 aligned (77%): 12 of 16 fitted + 21 of 27 unfitted.
% The 50 per cent control is 1,166 of 2,331 (R > 10 systems with a cataloged PA).

\end{abstract}
%\keywords{stars \& stellar, instrument \& method}

\section{Introduction} \label{sec:intro}
 
The third \gaia{} data release \citep[DR3;][]{GaiaDR3} marks the first time the \gaia{} mission \citep{GaiaMission} published large-scale astrometric solutions for non-single stars (NSS). For sources whose on-sky motion departs significantly from the standard five-parameter single-star model, the \gaia{} NSS pipeline fits higher-order astrometric solutions, including acceleration solutions in which the proper motion is allowed to vary linearly (\texttt{Acceleration7}) or quadratically (\texttt{Acceleration9}) with time \citep{Gaia_NSS}. Together with the full Keplerian orbital solutions \citep{Gaia_NSS,Holl2023}, the NSS solutions provide an all-sky census of unresolved binaries. The acceleration solutions are sensitive to companions with orbital periods comparable to or longer than the \gaia{} DR3 baseline ($\approx34$ months).

The seven-parameter solutions add a constant acceleration in right ascension (RA, $\alpha$) and declination (Dec, $\delta$) to the standard five-parameter model, while the nine-parameter solutions add a constant jerk in RA and Dec as well. Unlike for a Keplerian orbit, fitting a seven- or a nine-parameter sky path to measured positions is a linear problem, and the \gaia{} pipeline tries the nine-parameter model first and retains it only when the jerk is significant, falling back to the seven-parameter model otherwise \citep{Gaia_NSS}. The two families therefore probe different period regimes. For a circular orbit of period $P$ the jerk is $2\pi/P$ times the acceleration, so over a window of length $T \approx 2.8$~yr the cubic (jerk) term of the sky path reaches a fraction $\pi T/3P$ of the quadratic (acceleration) term at the edges of the window: about $0.1$ for $P = 30$~yr, $0.3$ for $10$~yr and unity for $3$~yr. A seven-parameter solution therefore describes an orbit with a period of tens of years, over which the acceleration is effectively constant, whereas a nine-parameter solution is assigned when the change in acceleration is detected, for periods of a few times $T$ or for an eccentric orbit passing periastron inside the window \citep{Gaia_NSS}.
 
Because these astrometric acceleration solutions are new, their reliability across the broad range of magnitudes, colors, and orbital configurations spanned by the catalog has not been fully characterized. Independent, physically grounded benchmarks are therefore valuable. Complementary acceleration measurements exist in the form of the \Hipparcos--\gaia{} proper motion anomaly \citep{Kervella2019} and the \Hipparcos--\gaia{} Catalog of Accelerations \citep[HGCA;][]{HGCA}, and recent work has validated \gaia{} NSS solutions against orbits derived from combined absolute astrometry and radial velocities \citep{156_systems}. Here we pursue a different and largely independent check: we ask whether the acceleration that \gaia{} reports for a star is quantitatively consistent with the gravitational pull of a \emph{known, resolved} companion.
 
Resolved wide binaries are well-suited to this test. The Washington Double Star catalog \citep[WDS;][]{WDS} tabulates component magnitudes, angular separations, and position angles for more than $10^5$ visual pairs. For a companion of mass $M_2$ at instantaneous separation $r$ from the primary, the gravitational acceleration imparted on the primary is $G M_2 / r^2$. The true separation is not observable, but its sky projection $s = r\sin\theta$ is, where $\theta$ is the angle between the separation vector and the line of sight. Because $s \le r$, evaluating the acceleration at $s$ gives a Newtonian upper bound on the acceleration, $G M_2 / s^2$, saturated when the separation vector lies in the plane of the sky. Divided by the distance, this bound becomes $\amax$, the maximum angular acceleration the companion can impart (Section~\ref{ssec:amax}). Because \gaia{} tracks the photocenter of a pair it does not resolve, the acceleration it measures is that of the primary reduced by a factor set by the flux ratio and the mass ratio, which we carry through both the bound and the orbit fits.

Comparing the \gaia{}-measured acceleration against this bound defines a diagnostic ratio $\aratio \equiv \ameas / \amax$ that tests whether the cataloged companion can account for the observed signal. For a genuine binary with the correct companion mass and a separation measured at the \gaia{} epoch, projection alone confines the ratio to $\aratio \le 1$, with a distribution over random orientations that we derive in Section~\ref{ssec:ratio}. Catalog resolution scatters $\aratio$ upward from this expectation, because a separation recorded at the resolution limit is an upper limit on $s$ and hence a lower limit on $\amax$. A pronounced excess ($\aratio \gg 1$) signals a closer, unseen companion---the wide pair alone cannot explain the measured acceleration. Systems with $\aratio \lesssim 1$ are consistent with the cataloged companion being responsible. $\aratio \ll 1$ is uncommon for a bound pair, requiring a companion that is far from its host physically yet close in projection. More likely such a value signals an error in the mass estimate, the WDS entry, or the acceleration solution.

In this paper we construct such a sample, quantify the acceleration ratio across the full cross-matched catalog, identify a clean subsample of physically plausible wide binaries for detailed follow-up, and test the direction of the acceleration against the companion position angle, a check that uses neither mass nor separation. For the plausible systems with enough resolved astrometry we attempt full orbital fits using the \orvara{} code \citep{orvara}, which we have extended to accept \gaia{} DR3 NSS acceleration and jerk coefficients as inputs, alone or jointly with the \Hipparcos--\gaia{} proper motion anomaly where a \Hipparcos{} counterpart exists. The updated packages are available at \url{https://github.com/t-brandt/orvara} and \url{https://github.com/gmbrandt/HTOF} (Section~\ref{ssec:orvara_ext}).
 
The paper is organized as follows. Section~\ref{sec:data} describes the \gaia{} DR3 NSS acceleration catalog, the WDS, and our cross-matching procedure. Section~\ref{sec:methods} details the estimation of primary and companion masses, the predicted maximum acceleration $\amax$, the acceleration ratio $\aratio$, the orbit-fitting setup, and the direction test. Section~\ref{sec:results} presents the full-sample statistics, the plausible wide-binary subsample and its orbit fits, the test of the acceleration direction, and the updates and extensions we have made to \orvara{} and \texttt{htof}. Section~\ref{sec:discussion} assesses the significance of the direction test, sets out the caveats that apply when using DR3 acceleration solutions, and considers prospects for \gaia{} DR4. Section~\ref{sec:conclusion} summarizes our findings.

\section{Data} \label{sec:data}

We begin by constructing a sample of stars that have both a \gaia{} DR3 acceleration solution and relative astrometry of a resolved companion. This section describes the two catalogs and their cross-match.
 
\subsection{Gaia DR3 Astrometric Accelerations} \label{ssec:gaia}
 
We query the \gaia{} DR3 \texttt{nss\_acceleration\_astro} table \citep{Gaia_NSS} for all sources with \texttt{Acceleration7} or \texttt{Acceleration9} solution types and non-null parallaxes, retaining the position, the parallax, the two acceleration components ($\dot{\mu}_{\alpha*}$, $\dot{\mu}_\delta$, in $\masyr$) with their uncertainties, the solution type, and \texttt{significance}, which yields \nnss acceleration solutions in total (\nacca \texttt{Acceleration7} and \naccb \texttt{Acceleration9}). The \texttt{Acceleration7} model adds the two acceleration components to the five-parameter single-star model, while \texttt{Acceleration9} additionally includes two jerk components ($\ddot{\mu}_{\alpha*}$, $\ddot{\mu}_\delta$), the time derivatives of the acceleration.

For an \texttt{Acceleration9} source the tabulated acceleration is the value at the reference epoch J2016.0 and the jerk its rate of change, so for companions with orbital periods comparable to the $\sim$34-month DR3 baseline the jerk terms absorb orbital curvature and the epoch-specific acceleration may differ appreciably from the orbit-averaged value. Such cases are present in the catalog by design: \citet{Gaia_NSS} retained acceleration solutions for many binaries with periods shorter than the mission duration, judging a polynomial fit safer than an orbit with an uncertain period. Because \gaia{} DR3 releases no epoch-level astrometry, none of its astrometric solutions can be independently refit, and this will remain so until epoch astrometry is released in DR4.

The DR3 data model defines \texttt{significance} as the norm of the highest-order vector of the model divided by its uncertainty, the acceleration for \texttt{Acceleration7} and the jerk for \texttt{Acceleration9} \citep{GaiaDR3doc}. The uncertainty is taken from a covariance matrix whose per-transit errors have been rescaled to force the corrected goodness-of-fit $F_2 = 0$. It is an amplitude signal-to-noise of the highest-order fitted term, not a comparison between nested models. We use \texttt{significance} quantitatively in two places: to compare it against a direct signal-to-noise on the acceleration amplitude (Section~\ref{ssec:ratio}), and to identify the system with the least stable acceleration across the DR3 baseline (02136$-$0849, Section~\ref{ssec:orbitfit_results}).
 
\subsection{Washington Double Star Catalog} \label{ssec:wds}
 
The Washington Double Star (WDS) catalog \citep{WDS} is the principal database of resolved double and multiple star systems, containing positional measurements, component magnitudes, and discovery information for over 150,000 systems. We use the version of the summary catalog current as of 2026 June, with \nwdscat entries.
 
The summary catalog carries one line per pair: the J2000 coordinates of the primary, the discoverer designation and, for a resolved multiple, the component label (e.g., \texttt{Aa,Ab}), the epochs of the first and last observation with the position angle and angular separation at each (PA$_1$, $\rho_1$, PA$_2$, $\rho_2$), the number of observations, the component magnitudes ($m_1$, $m_2$), and a notes field whose flags include the photometric band of the magnitudes. The epochs, angles, separations, and count are drawn only from measures that record all three of date, position angle, and separation, and a pair with no such measure has these fields set to $-1$. We use these summary fields for the cross-match and for the population-level analysis. The magnitudes are predominantly in the visual ($V$) band (91\% of the matched sample carry no red, blue, or infrared flag) but are heterogeneous, and the close pairs that make up the plausible sample are more often measured in red or infrared bands, which has consequences for the inferred mass ratios that we discuss in Section~\ref{sssec:m2_sys}. Separations are recorded to $0\farcs1$, so a cataloged $\rho_1 = 0\farcs1$ is an upper limit rather than a measurement (Section~\ref{ssec:amax}). 

For the plausible systems of Section~\ref{ssec:ratio} and the two systems just outside the window, 16206$+$4535 and 18267$-$3024, we additionally obtained the WDS measure files by data request to the U.S.\ Naval Observatory. A measure file lists every individual measurement of a pair with its epoch, position angle and separation. Two kinds of entry are flagged and are not used here: occultation vector angles, which record the orientation of the occulting limb rather than the position angle of the pair, and measures the WDS marks as erroneous. The remaining measures supply the relative astrometry for the orbit fits (Section~\ref{ssec:orbitfit}) and the epoch-matched position angles for the direction test (Section~\ref{ssec:direction}).

\subsection{Cross-Matching Procedure} \label{ssec:xmatch}

We cross-match the \nnss NSS acceleration solutions of Section~\ref{ssec:gaia} against the Washington Double Star catalog \citep[WDS; \nwdscat entries;][]{WDS} in three steps. First, because the WDS tabulates positions at J2000.0 while the DR3 positions refer to J2016.0, we propagate every NSS source to J2000.0 with its DR3 proper motion. Second, each propagated source is matched to its nearest WDS entry on the sky, and we retain only pairs with angular separations $<2\arcsec$. Third, where several \gaia{} sources fall within the match radius of the same WDS entry, we keep only the closest, so that each WDS system enters the sample once.

The epoch propagation is essential for the high-proper-motion pairs. A proper motion of $125\,{\rm mas\,yr^{-1}}$ accumulates to $2\arcsec$ over the 16-yr epoch difference, equal to the match radius, so without the propagation any faster-moving source would fail to match. The propagation recovers 278 matches and removes 4, and 254 of the matched sources have $\mu > 125\,{\rm mas\,yr^{-1}}$. The cross-match yields \nmatch positional matches within $2\arcsec$, \nuniq unique WDS entries after de-duplication, and \nmagpair of these (\pmagpair{}\%) with both component magnitudes $m_1$ and $m_2$ cataloged.
 
The $2\arcsec$ match radius and the heterogeneous nature of the WDS introduce two forms of contamination: chance coincidences between an NSS source and an unrelated WDS entry, and genuine WDS pairs that are optical doubles rather than bound systems. The random-match rate is small. Taking the $N_{\rm WDS} = \nwdscat$ entries to be distributed uniformly over the sky, the probability that a given NSS source has an unrelated entry within the match radius $r_{\rm m} = 2\arcsec$ is 
\begin{equation} \label{eq:chance} 
p \simeq \frac{N_{\rm WDS}\,\pi r_{\rm m}^{2}}{4\pi} = \frac{N_{\rm WDS}\,r_{\rm m}^{2}}{4}\simeq 4\times10^{-6} 
\end{equation} 
($r_{\rm m}$ in radians), or $\sim$1 random match among the \nnss NSS sources. This uniform-density estimate is a floor---both catalogs cluster strongly on the sky, enhancing the local rate---but the same selection effects make them sample the same population of bright, nearby, often-multiple stars, so the \nuniq matches are overwhelmingly physical rather than random superpositions. The more relevant channel is therefore the optical double, in which the WDS pairs the NSS primary with an unbound foreground or background star. For such a pair the cataloged separation and photometric mass yield an $\amax$ (Section~\ref{ssec:amax}) that has no physical meaning. Both channels are suppressed by the acceleration-ratio selection of Section~\ref{ssec:ratio}, where we return to them once the ratio is defined.
 
\section{Methods} \label{sec:methods}

Our next step is to downselect the cross-matched catalog to physically bound systems for which the \gaia{} NSS solution represents motion due to the resolved, cataloged companion. For each matched pair we therefore need the largest acceleration that companion could impart to the \gaia{}-tracked photocenter. This Newtonian upper bound, $\amax$, depends on the companion mass ($M_2$), the distance to the primary as inferred from the \gaia{} parallax ($d$), and the projected physical separation of the two components ($s = \rho_1 d$, where $\rho_1$ is the WDS first-epoch angular separation). It also depends on the offset of the photocenter from the barycenter, which is set by the flux ratio and the mass ratio of the pair. The bound is the gravitational acceleration that $M_2$ would exert on the primary at the projected separation, $GM_2/s^2$, divided by $d$ to convert it to an angular acceleration and multiplied by a photocenter factor that accounts for this offset. Neither component mass is measured directly. We therefore estimate the primary mass ($M_1$) first and then derive $M_2$ from $M_1$ and the WDS magnitude difference between the two components. Comparing the measured acceleration $\ameas$ with $\amax$ gives one ratio per system, $\aratio = \ameas/\amax$. Systems in which the two agree to within an order of magnitude form the plausible sample of Section~\ref{ssec:ratio}, from which the orbit-fitting targets are drawn.

\subsection{Primary Mass Estimation} \label{ssec:m1}

The WDS magnitude difference constrains the companion mass only relative to the primary, through the mass--luminosity relation that we discuss in the following section. The bound $\amax$ therefore rests on the primary mass, which we estimate in three tiers. The first uses the \gaia{} source itself, while the second and third cross-match against external catalogs on positions propagated from J2016.0 to the catalog epochs, J1991.25 for \Hipparcos{} and J2000.0 for 2MASS, with the DR3 proper motions as in Section~\ref{ssec:xmatch}.
 
\begin{enumerate}
    \item \textbf{\gaia{} DR3 \texttt{mass\_flame}}: We take stellar masses derived from the FLAME module of the \gaia{} DR3 astrophysical parameters \citep{GaiaDR3_overview, Fouesneau2023}. This provides $M_1$ for \nflame of \nuniq sources (\pflame{}\%).
 
    \item \textbf{\Hipparcos{} fallback}: For sources without a \gaia{} \texttt{mass\_flame} value, we cross-match against the \Hipparcos{} catalog \citep{Hipparcos_1997} at $2\arcsec$ to obtain a \textsc{hip} identifier, and adopt the corresponding stellar mass from the proper-motion-anomaly catalog of \citet{Kervella2019}. This provides $M_1$ for a further \nhipmass sources (\phipmass{}\%).

    \item \textbf{2MASS $J$-band fallback}: For the remaining sources, we cross-match with the 2MASS Point Source Catalog \citep{2MASS} at $2\arcsec$ and convert the absolute magnitude $M_J = J - 5\log_{10}(d/10\,\mathrm{pc})$, with $d = 1000\,\mathrm{pc}/(\varpi/\mathrm{mas})$ from the \gaia{} parallax $\varpi$, to a mass by interpolating $\log_{10} M_1$ linearly in $M_J$ on the main-sequence table of \citet[][version 2022.04.16]{Pecaut2013}, which runs from O9V to L2V and follows the $M_J$ relation of \citet{Delfosse2000} and the $M_{K_s}$ relation of \citet{Mann2019} to within 10\% for M dwarfs. This provides $M_1$ for \ntwomass of the \ntwomassin sources entering this tier (\ptwomass{}\% of the full sample), and the \nnomone without one are brighter than the O9V end of the table or lack a $J$ magnitude.
\end{enumerate}
 
Together these three tiers yield a primary-mass estimate for \nmone of the \nuniq systems (\pmone{}\%). The remaining \nnomone (\pnomone{}\%) lack a usable $M_1$ from any tier, and because $M_2$ is derived from $\Delta m$ relative to $M_1$, they have no companion mass and no ratio, so they are excluded from the acceleration-ratio analysis.

The three tiers are not on a common mass scale. For the 26 plausible systems with a FLAME mass, the FLAME value lies 8\% below the 2MASS-$J$ estimate on median, and for the 14 with both FLAME and \Hipparcos{} masses the FLAME value lies 9\% below the \Hipparcos{} one. Because $M_1$ enters $\amax$ linearly through $M_2 = qM_1$, offsets of this size propagate directly into $\aratio$, where they are subdominant to the order-of-magnitude width of the plausible window and to the separation systematic of Section~\ref{ssec:amax}.

\subsection{Companion Mass Estimation} \label{ssec:m2}

For each system with a primary mass $M_1$ and a magnitude difference between the components, we estimate the companion mass $M_2$ by inverting a mass--luminosity relation. This subsection gives the inversion, the sources of the magnitude difference, and the systematic uncertainties of the result.

\subsubsection{Mass--Luminosity Inversion} \label{sssec:m2_ml}

The magnitude difference $\Delta m = m_2 - m_1$ yields a flux ratio in the band of the magnitudes, predominantly $V$ (Section~\ref{ssec:wds}),
\begin{equation}
    \phi \equiv \frac{f_2}{f_1} = 10^{-0.4 \Delta m}.
\end{equation}
We take this band flux ratio as the bolometric luminosity ratio of the two components, which is exact only when the components have the same bolometric correction, and adopt a bolometric mass--luminosity relation of the form $L \propto M^{\gamma}$, so that $\phi \approx L_2/L_1 = (M_2/M_1)^{\gamma}$. The companion mass is then the root of
\begin{equation} \label{eq:m2_root}
    f(M_2) = \left(\frac{M_2}{M_1}\right)^{\gamma(M_2)} - \phi = 0,
\end{equation}
where the exponent $\gamma$ is piecewise in mass and is evaluated at the companion mass in solar units,
\begin{equation} \label{eq:gamma}
    \gamma(M_2) = \begin{cases}
        2.3 & M_2 < 0.43 \\
        4.0 & 0.43 \leq M_2 < 2.0 \\
        3.5 & M_2 \geq 2.0.
    \end{cases}
\end{equation}
The break points and exponents are the standard piecewise bolometric mass--luminosity relation \citep{SalarisCassisi2005, Eker2018}, with $\gamma \approx 2.3$ below $0.43\,\Msun$, $\gamma \approx 4$ across the lower main sequence, and $\gamma \approx 3.5$ for intermediate-mass stars, applied here to a flux ratio measured in $V$ or a redder band. In $V$ the relation is steeper at the M-dwarf end \citep{Delfosse2000}, so Equation~\ref{eq:gamma} underestimates the masses of M-dwarf companions, a systematic we quantify in Section~\ref{sssec:m2_sys}. The root is found numerically, so that $\gamma$ is evaluated at the solution $M_2$ rather than at $M_1$. In the rare case where the bracket $[0.01, 20]\,\Msun$ does not contain a root, we fall back to the closed-form estimate $M_2 = M_1\,\phi^{1/\gamma(M_1)}$, which holds the exponent at the primary's regime and is exact when both components share the same segment of Equation~\ref{eq:gamma}. Two systems in the full matched sample, 01292$+$4820 and 01494$-$1042, both with $\aratio \gg 10$, required this path, and none of the plausible systems did.

\subsubsection{Sources of the Magnitude Difference} \label{sssec:m2_src}

The magnitude difference comes from the WDS component magnitudes or, when they are not available, from an assumed mass ratio in place of any photometry. For \ndmwds systems (\pdmwds{}\%) it is the WDS magnitude difference $\Delta m = m_2 - m_1$ of Section~\ref{ssec:xmatch}. The \nmissmtwo systems (\pdmnone{}\%) with a valid primary but no cataloged secondary magnitude are pairs at $0\farcs2$--$5\arcsec$ for which \gaia{} DR3 resolves no companion sharing the primary's parallax, so no $\Delta m$ can be recovered from \gaia{} either. These receive $M_2 = q_{\rm typ}\,M_1$, where $q_{\rm typ}$ is a typical wide-binary mass ratio that depends on the primary mass: $q_{\rm typ} = 0.5$ for $M_1 \leq 0.8\,\Msun$, $0.3$ for $0.8 < M_1 \leq 1.5\,\Msun$, and $0.4$ for $M_1 > 1.5\,\Msun$ \citep{Duchene2013, Moe2017}. For these the flux ratio entering the photocenter factor of Section~\ref{ssec:amax} is the one implied by $q_{\rm typ}$ through Equation~\ref{eq:gamma}.

The root-finder of Section~\ref{sssec:m2_ml} provides $M_2$ for the \nmtworoot systems (\pmtworoot{}\% of the \nuniq matched sources) that have both an $M_1$ and a WDS magnitude difference, and the mass-ratio fallback covers the remaining \ndmnone, so that every one of the \nmone systems with a primary mass (Section~\ref{ssec:m1}) is assigned an $M_2$. Of the \nplaus systems in the plausible subsample (Section~\ref{ssec:ratio}), 42 derive $M_2$ from the WDS magnitude difference and three (08358$+$0637, 14460$+$4633 and 22526$+$5643, marked in Table~\ref{tab:plausible}) have no secondary photometry and receive the mass-ratio fallback. For these three, $M_2$, $\fpc$ and hence $\aratio$ rest on the assumed $q_{\rm typ}$ rather than on a measurement, and they enter the population statistics with that caveat. None of the \nplaus required the closed-form path.

\subsubsection{Systematic Uncertainties in the Companion Mass} \label{sssec:m2_sys}

The dominant systematic in the magnitude-based $M_2$ is the heterogeneous photometry underlying the WDS magnitude difference: the WDS is not a photometric catalog, and the flux ratio is not always in the band the relation assumes. The WDS \texttt{Notes} flags identify the band of the cataloged magnitudes, and a substantial fraction of the sample reports them in a band redder than $V$. Of the \nplaus plausible systems, \nredflag carry the \texttt{R} flag (red, Johnson $R/I$ or red photographic) and \nkflag carry the \texttt{K} flag (near-infrared $K$ band), \nbandflag in total. Because the pipeline forms $\Delta m = m_2 - m_1$ without a band correction, applying the relation to a red or infrared flux ratio shifts the inferred $M_2$ by an amount set by the color difference between the two components. A second, related systematic is the differential bolometric correction: the bolometric correction in $V$ grows from near zero for FGK dwarfs to $1$--$2$~mag for mid-M dwarfs, so a $V$-band flux ratio understates the luminosity ratio of an M-dwarf companion to an FGK primary. Equivalently, for companions below $0.43\,\Msun$ the bolometric $\gamma = 2.3$ of Equation~\ref{eq:gamma} is shallower than the $V$-band relation \citep{Delfosse2000}, and $M_2$ is biased low for M-dwarf companions by up to a factor of $\sim$1.7 at $\Delta m \approx 3$~mag. The two effects act in the same direction on $\amax$ and are not separable with the WDS photometry alone. Both propagate into $\amax$ only through the companion-mass solve and remain subdominant to the separation systematic (Section~\ref{ssec:amax}).

\nswapped systems in the full matched sample exhibit $\Delta m < 0$, the cataloged ``secondary'' being brighter than the ``primary'', which indicates swapped component labels. The root-finder then returns $M_2 > M_1$ ($q > 1$), biasing $\amax$ for those systems, but none of them survive into the \nplaus plausible binaries. A further \nequal systems have equal cataloged magnitudes ($\Delta m = 0 \Rightarrow M_2 = M_1$), for which the photocenter correction of Section~\ref{ssec:amax} gives $\amax = 0$, and they are set aside as untestable. Both groups are retained and flagged in the matched sample rather than removed, so they are counted in the totals quoted above, and neither affects the population-level conclusions.

\subsection{Predicted Maximum Acceleration} \label{ssec:amax}
For each binary system, we compute the maximum angular acceleration that the WDS companion of mass $M_2$ could produce on the primary if the separation vector lies in the plane of the sky:
\begin{equation} \label{eq:alpha_max}
    \amax = \frac{G\,M_2}{s^2\,d},
\end{equation}
where $s = \rho_1\,d$ is the projected physical separation, $\rho_1$ is the WDS first-observation angular separation (the \texttt{sep1} field), and $d$ is the distance from the parallax (Section~\ref{ssec:m1}), so that $s$ is in AU when $\rho_1$ is in arcsec and $d$ in pc. The numerator $G M_2 / s^2$ is the linear acceleration of the primary, and dividing by $d$ converts it to an angular acceleration under the small-angle approximation. We adopt the companion mass $M_2$ (rather than the total mass) because the astrometric acceleration of the primary is set by the gravitational pull of the companion alone, $a_1 = G M_2 / s^2$.

\gaia{}, however, does not track the primary: for a pair unresolved by the NSS pipeline it tracks the photocenter, which lies a fraction $\beta = \phi/(1+\phi)$ of the way from the primary to the secondary, whereas the barycenter lies at $B = M_2/(M_1+M_2) = q/(1+q)$. Both points move on Keplerian orbits about the barycenter with the same period, and the photocenter's orbit is the primary's scaled by $(B-\beta)/B$, so the acceleration that \gaia{} can measure is
\begin{equation} \label{eq:photocenter}
    \amax = \left|1 - \frac{\beta}{B}\right| \frac{G\,M_2}{s^2\,d},
\end{equation}
with $\phi$ the flux ratio of Section~\ref{ssec:m2}. For a faint companion $\beta \to 0$ and Equation~\ref{eq:photocenter} reduces to Equation~\ref{eq:alpha_max}. For a pair of equal brightness $\beta = B = 1/2$ and the photocenter does not accelerate at all. This photocenter factor, which we denote $\fpc \equiv |1-\beta/B|$, has median $\fpcfull$ over the full matched sample and $\fpcplaus$ over the plausible sample and is tabulated for each plausible system in Table~\ref{tab:plausible}. The \nequal equal-magnitude pairs ($\amax = 0$) are excluded from the ratio statistics. All values of $\amax$ and $\aratio$ quoted in this paper, and those in Table~\ref{tab:plausible}, use the photocenter-corrected bound of Equation~\ref{eq:photocenter}. Equation~\ref{eq:alpha_max} is retained only to define the limit of a faint companion. Without the correction the plausible window would admit \nuncorr rather than \nplaus systems. The correction removes twelve equal-magnitude pairs and nine pairs whose $\aratio$ it pushes above 10, and admits one pair, 08358$+$0637, whose uncorrected $\aratio = 0.09$ it raises to $\minR$.
 
We use the first-epoch separation $\rho_1$ as the fiducial projected separation. With a measured $\rho_1$, Equation~\ref{eq:photocenter} is an upper limit on the measurable acceleration, as argued in Section~\ref{sec:intro} and quantified in Section~\ref{ssec:ratio}. Two properties of the catalog limit how far $\rho_1$ can be trusted as a measurement. For systems with significant orbital motion between the first and last WDS epochs, $\rho_1$ and the last-epoch separation $\rho_2$ can differ, and we treat this difference as part of the separation systematic. More importantly, at the catalog resolution limit $\rho_1$ is not a measurement at all: \nrhoquant of the \nplaus plausible systems have $\rho_1 = 0.1\arcsec$, the smallest separation the speckle and occultation measures dominating this subsample report, so for these the cataloged value is better read as an upper bound on the projected separation. Since $\amax \propto \rho_1^{-2}$, an overstated $\rho_1$ makes the computed $\amax$ too small and the computed $\aratio$ too large, so for these systems $\aratio$ overestimates the true ratio.

This is the most likely origin of the median $\aratio = \medR$ of the plausible sample lying above unity. The isotropic expectation of Section~\ref{ssec:ratio} makes the excess quantitative: only $\fRbelowone{}\%$ of the plausible sample has $\aratio < 1$ and $\fRbelowmed{}\%$ has $\aratio < 0.65$, against $100\%$ and $50\%$ expected, and the observed median is $\medRiso$ times the predicted $0.65$. Since $\aratio \propto \rho_1^{2}$ at fixed $\ameas$, this corresponds to a typical over-estimate of the separation by a factor of about two, consistent with the reduction in scatter obtained in Section~\ref{ssec:masses} when $\rho_1$ is replaced by the separation implied by the fitted orbit and the photocenter by the primary.
 
\subsection{Acceleration Ratio} \label{ssec:ratio}
 
The measured sky-plane acceleration is computed from the \gaia{} NSS acceleration coefficients as
\begin{equation} \label{eq:alpha_meas}
    \ameas = \sqrt{\alpha_{\rm RA}^2 + \alpha_{\rm Dec}^2},
\end{equation}
where $\alpha_{\rm RA}$ and $\alpha_{\rm Dec}$ are the time derivatives of the proper-motion components $\mu_{\alpha*}$ and $\mu_\delta$, reported in mas\,yr$^{-2}$ as \texttt{accel\_ra} and \texttt{accel\_dec} \citep{Gaia_NSS}. For \texttt{Acceleration7} solutions these are constant over the \gaia{} baseline. For \texttt{Acceleration9} solutions a jerk term is also fitted, so the modeled acceleration varies linearly with time. The tabulated coefficients then give its value at the DR3 reference epoch J2016.0, and we adopt that value as $\ameas$, noting that for periods comparable to the DR3 baseline it can differ appreciably from the orbit-averaged acceleration (Section~\ref{ssec:gaia}). The direction test of Section~\ref{ssec:direction} uses the position angle of the same vector, $\mathrm{PA}_{\rm acc} = \mathrm{atan2}(\alpha_{\rm RA}, \alpha_{\rm Dec})$, measured from north through east.

We define the acceleration ratio as
\begin{equation} \label{eq:ratio}
    \aratio \equiv \frac{\ameas}{\amax},
\end{equation}
which compares the observed total sky-plane acceleration to the maximum Newtonian prediction from the known WDS companion. A ratio $\aratio \lesssim 1$ indicates consistency with the companion being the sole source of gravitational perturbation. Ratios $\aratio \gg 1$ suggest additional unseen companions (such as unresolved inner binaries driving the NSS signal) or systematic effects in the acceleration solution. We define the ``physically plausible'' range as $0.1 \leq \aratio \leq 10$, encompassing systems where the measured acceleration is within one order of magnitude of the Newtonian prediction. For a genuine binary the ratio is bounded by projection. The true acceleration of the primary is $GM_2/r^2$, of which \gaia{} measures only the sky-plane component, $(GM_2/r^2)\sin\theta$. The bound $\amax$ is instead evaluated at the projected separation $s = r\sin\theta$, which makes it larger than the true acceleration by a factor $1/\sin^2\theta$. The ratio of the two is $\aratio = \sin^3\theta$, which holds at any instant of any orbit, so $\aratio \le 1$. For randomly oriented systems $\cos\theta$ is uniformly distributed, so that $P(\aratio < x) = 1 - \sqrt{1 - x^{2/3}}$. This distribution has a median of $(3/4)^{3/2} = 0.65$ and a mean of $3\pi/16 = 0.59$, and places $11\%$ of orientations below $\aratio = 0.1$. The lower edge of the plausible window therefore excludes about a tenth of genuine binaries, while the upper edge admits a factor of ten of separation-driven excess above the physical bound.
 
Of the \nuniq matched systems, \nfinite have a finite and physically meaningful acceleration ratio. The other \nnoratio are excluded because they lack a primary mass (\nnomone), have no cataloged first-epoch separation (\texttt{sep1} zero or absent, \nsepzero), or are equal-magnitude pairs for which $\amax = 0$ (\nequal, five of which also lack a separation). The median ratio of the \nfinite is $\aratio \approx \medRfull\times10^{5}$, and the overwhelming majority exceed the wide-companion prediction by more than an order of magnitude ($\aratio > 10$). These positional matches are either optical pairs or hierarchical systems in which the \gaia{} NSS acceleration is driven by an unresolved inner companion rather than the cataloged wide WDS pair.
 
Imposing $0.1 \leq \aratio \leq 10$ isolates \nplaus systems, which constitute our plausible sample. These span \nplausa \texttt{Acceleration7} and \nplausb \texttt{Acceleration9} solutions, 42 of them with companion masses derived from the WDS magnitude difference and three (08358$+$0637, 14460$+$4633 and 22526$+$5643) from the assumed mass ratio of Section~\ref{sssec:m2_src}. Their ratios range from $\minR$ to $\maxR$ with a median of $\medR$ (mean $\meanR$), and for \nexplain of the \nplaus the measured acceleration is at least half the face-on Newtonian maximum ($\aratio \geq 0.5$). \Nplauship of the \nplaus have \Hipparcos{} counterparts (Section~\ref{sec:results}). Two systems fall just outside the window, 18267$-$3024 (HIP 90397) below it at $\aratio = 0.08$ and 16206$+$4535 above it at $\aratio = \Rboundary$ ($\Rboundaryunc$ without the photocenter correction). 18267$-$3024 is an \texttt{Acceleration9} source whose companion passes through conjunction inside the DR3 window, so the cataloged acceleration is a window average across a reversal of direction, an order of magnitude below the value the WDS separation predicts, and its jerk is $5.9$~yr$^{-1}$ times its acceleration. A Mode~B fit (Section~\ref{ssec:orbitfit}) to its three resolved epochs converges but reproduces neither the relative astrometry ($\chi^2_\rho = 49$ for five measures) nor the direction of the acceleration, and returns an unphysical mass ratio, as expected when the sky path is strongly curved inside the window (Section~\ref{ssec:gaia}), so we report it here and not in Tables~\ref{tab:orbits} and \ref{tab:chisq}. 16206$+$4535 carries the best-sampled resolved orbit in the matched catalog, and we retain it as a labeled boundary case in the orbit fits of Section~\ref{ssec:orbitfit_results}, which test each system on its own, but not in the population statistics, which are defined on the sample selected by the window.

Because the cataloged \texttt{significance} does not measure the acceleration amplitude for \texttt{Acceleration9} sources (Section~\ref{ssec:gaia}), we form a direct signal-to-noise ratio on $\ameas$. We propagate \texttt{accel\_ra\_error} and \texttt{accel\_dec\_error} to first order through Equation~\ref{eq:alpha_meas}, $\sigma_\alpha = \sqrt{(\alpha_{\rm RA}\sigma_{\rm RA})^2 + (\alpha_{\rm Dec}\sigma_{\rm Dec})^2}/\ameas$, neglecting the correlation between the two components, and define ${\rm S/N} = \ameas/\sigma_\alpha$. The Pearson correlation coefficient between this S/N and \texttt{significance} across the plausible sample is \pearsona for the \nplausa \texttt{Acceleration7} systems, for which the two quantities measure the same term, and \pearsonb for the \nplausb \texttt{Acceleration9} systems, for which \texttt{significance} refers to the jerk. Every acceleration in the plausible sample is detected at S/N $\ge 17.8$, the lowest two values belonging to the \texttt{Acceleration7} solutions of 00350$-$6342 (S/N $= 17.8$) and 13097$-$2212 ($18.5$), and all \texttt{Acceleration9} accelerations are detected at S/N $\ge 20$. The uncertainty on $\ameas$ is thus at most $6\%$ for any plausible system, so the scatter in $\aratio$ is set by $\amax$ and not by the measured acceleration.

The acceleration-ratio selection also suppresses both contamination channels identified in Section~\ref{ssec:xmatch}: an optical pair mistaken for a physical one, and a physical wide pair whose \gaia{} acceleration is produced by a different body. An unrelated companion exerts negligible pull on the primary, so a real \gaia{} acceleration must come from another perturber, typically a closer, unresolved companion. The $\amax$ predicted from the cataloged wide component is then unrelated to the measured value and usually far below it, which places the system in the $\aratio \gg 1$ tail outside the window. Contaminating the final \nplaus-system sample thus requires an unrelated star whose inferred $\amax$ matches the measured acceleration to within an order of magnitude by chance. Such cases cannot be excluded from photometry and projected separation alone, but they are tested per system by the orbit fits of Section~\ref{ssec:orbitfit}, which combine the WDS relative astrometry with the \gaia{} acceleration and, where available, the \Hipparcos{} astrometry, and by the acceleration--companion direction test of Section~\ref{ssec:direction}, both of which require consistency beyond amplitude agreement.

%\startlongtable
\begin{deluxetable*}{lccccccccccc}
\tabletypesize{\scriptsize}
\tablecaption{The \nplaus plausible wide-binary systems ($0.1 \le \aratio \le 10$) and the boundary case 16206$+$4535.
\label{tab:plausible}}
\tablehead{
\colhead{WDS} & \colhead{Discoverer} & \colhead{HIP} & \colhead{Sol.} & \colhead{$\rho_1$} & \colhead{$M_1$} & \colhead{$M_1$ Ref.$^{a}$} & \colhead{$M_2$} & \colhead{$f_{\rm pc}$} & \colhead{$\amax$} & \colhead{$\ameas$} & \colhead{$\aratio$} \\[-0.6em]
\colhead{} & \colhead{} & \colhead{} & \colhead{} & \colhead{(\arcsec)} & \colhead{($\Msun$)} & \colhead{} & \colhead{($\Msun$)} & \colhead{} & \colhead{($\masyr$)} & \colhead{($\masyr$)} & \colhead{}
}
\startdata
00350-6342 & FR 3 &  & Acc7 & 0.5 & 0.79 & J & 0.13 & 0.89 & 0.783 & 1.11 & 1.42 \\
02136-0849 & TOK 361 & 10365 & Acc9 & 0.1 & 0.87 & G & 0.80 & 0.14 & 3.27 & 9.42 & 2.88 \\
03309-6200 & TOK 190 & 16370 & Acc9 & 0.1 & 1.29 & G & 0.91 & 0.52 & 16.3 & 3.59 & 0.22 \\
04248+1552 & PAT 7 GH &  & Acc9 & 0.2 & 1.06 & G & 0.25 & 0.81 & 2.14 & 15.4 & 7.21 \\
04407-2812 & TOK 209 &  & Acc9 & 0.2 & 0.96 & G & 0.26 & 0.78 & 3.12 & 8.55 & 2.74 \\
04433+1704 & PAT 14 &  & Acc9 & 0.1 & 0.79 & J & 0.46 & 0.72 & 12.6 & 16.0 & 1.27 \\
04476+1816 & PAT 17 & 22265 & Acc7 & 0.2 & 1.04 & H & 0.22 & 0.85 & 1.63 & 2.88 & 1.76 \\
05231-3142 & GKM9121 Aa,Ab & 25176 & Acc9 & 0.1 & 0.93 & H & 0.76 & 0.32 & 1.62 & 8.37 & 5.17 \\
05369-4758 & VSP 14 Ba,Bb & 26369 & Acc7 & 0.3 & 0.69 & H & 0.17 & 0.81 & 4.01 & 2.86 & 0.71 \\
06119-3529 & VSP 17 &  & Acc9 & 0.1 & 0.88 & J & 0.61 & 0.54 & 2.55 & 7.55 & 2.96 \\
08358+0637 & VI 3 Ba,Bb & 42173 & Acc7 & 0.2 & 0.87 & G & 0.26\tablenotemark{\small b} & 0.74 & 11.2 & 1.35 & 0.12 \\
08423+1322 & OCC9145 &  & Acc9 & 0.1 & 1.40 & G & 1.16 & 0.29 & 0.291 & 1.63 & 5.62 \\
08588-5303 & VSP 24 &  & Acc7 & 0.1 & 0.72 & J & 0.18 & 0.81 & 3.32 & 4.09 & 1.23 \\
09293-4432 & EGN 11 Ba,Bb & 46535 & Acc7 & 0.4 & 1.16 & G & 0.31 & 0.78 & 0.842 & 2.57 & 3.06 \\
10095-3251 & VSP 27 & 49767 & Acc7 & 0.3 & 1.07 & G & 0.07 & 0.97 & 0.353 & 1.55 & 4.38 \\
11181+1533 & GKM9551 Da,Dd &  & Acc7 & 0.1 & 0.75 & G & 0.47 & 0.65 & 5.68 & 4.95 & 0.87 \\
11253-8457 & BRC 4 Aa,Ab & 55746 & Acc7 & 0.1 & 1.40 & G & 1.25 & 0.18 & 1.35 & 5.42 & 4.03 \\
11585-7754 & KOH 91 & 58400 & Acc7 & 0.1 & 0.97 & G & 0.83 & 0.25 & 0.748 & 4.19 & 5.60 \\
12104-4352 & TOK 897 & 59341 & Acc7 & 0.1 & 1.09 & G & 0.90 & 0.29 & 0.575 & 4.33 & 7.53 \\
12146-1517 & GKM9281 Aa,Ab &  & Acc7 & 0.1 & 0.77 & G & 0.29 & 0.64 & 0.853 & 3.07 & 3.60 \\
13097-2212 & TOK 28 &  & Acc7 & 0.3 & 0.94 & J & 0.18 & 0.86 & 3.81 & 3.81 & 1.00 \\
14115-1204 & GKM9310 Aa,Ab &  & Acc7 & 0.1 & 0.87 & G & 0.26 & 0.74 & 0.810 & 3.11 & 3.85 \\
14382+1402 & TOK 406 & 71572 & Acc9 & 0.1 & 1.12 & H & 0.79 & 0.52 & 11.3 & 19.8 & 1.76 \\
14460+4633 & UYA 1 &  & Acc7 & 0.2 & 0.17 & J & 0.09\tablenotemark{\small b} & 0.49 & 8.26 & 6.02 & 0.73 \\
15210-1533 & OCC9209 Aa,Ab & 75118 & Acc7 & 0.1 & 1.59 & G & 0.76 & 0.85 & 14.6 & 5.22 & 0.36 \\
16077-2125 & MCA 43 & 79020 & Acc7 & 0.1 & 2.00 & G & 1.59 & 0.36 & 0.751 & 2.47 & 3.30 \\
16206+4535\tablenotemark{\small c} & HDS2309 & 80074 & Acc9 & 0.2 & 1.27 & G & 0.65 & 0.81 & 0.870 & 10.7 & 12.3 \\
16238-0258 & TOK 901 & 80315 & Acc7 & 0.2 & 0.87 & H & 0.18 & 0.85 & 2.15 & 2.46 & 1.14 \\
16252-3048 & USD1464 &  & Acc9 & 0.1 & 1.24 & G & 0.82 & 0.60 & 1.27 & 6.57 & 5.17 \\
16268-4936 & FR 18 & 80559 & Acc7 & 0.2 & 0.63 & H & 0.14 & 0.84 & 7.87 & 2.00 & 0.25 \\
16334-2015 & GKM9363 Aa,Ab &  & Acc7 & 0.2 & 0.80 & G & 0.25 & 0.73 & 0.912 & 2.41 & 2.65 \\
16387-3605 & GKM9364 Ba,Bb &  & Acc7 & 0.1 & 0.75 & J & 0.64 & 0.25 & 1.60 & 3.28 & 2.05 \\
18027-5714 & GKM9387 Aa,Ab &  & Acc7 & 0.1 & 0.80 & G & 0.19 & 0.82 & 3.92 & 3.10 & 0.79 \\
18060-4445 & GKM9390 Aa,Ab &  & Acc7 & 0.1 & 1.00 & J & 0.37 & 0.66 & 1.22 & 3.83 & 3.14 \\
19156-0830 & GKM9417 Aa,Ab &  & Acc7 & 0.2 & 0.71 & J & 0.23 & 0.71 & 0.547 & 3.51 & 6.41 \\
19169-4348 & GKM9420 Aa,Ab &  & Acc9 & 0.1 & 0.73 & G & 0.60 & 0.29 & 2.77 & 16.9 & 6.10 \\
20192-2610 & GKM9450 Ba,Bb &  & Acc7 & 0.1 & 0.70 & J & 0.26 & 0.66 & 0.693 & 2.61 & 3.77 \\
20335-1333 & OCC1089 &  & Acc7 & 0.1 & 1.23 & G & 1.00 & 0.32 & 0.237 & 1.77 & 7.47 \\
20467+1846 & YR 28 &  & Acc9 & 0.1 & 1.61 & G & 1.16 & 0.49 & 0.634 & 4.42 & 6.98 \\
22006-1410 & TOK 215 & 108649 & Acc7 & 0.2 & 1.24 & G & 0.33 & 0.78 & 0.797 & 3.70 & 4.65 \\
22084-3259 & VSP 56 & 109285 & Acc9 & 0.1 & 2.20 & H & 1.30 & 0.71 & 40.4 & 37.5 & 0.93 \\
22246+3926 & RAO 28 Aa,Ab & 110626 & Acc7 & 0.2 & 0.96 & G & 0.28 & 0.76 & 2.02 & 2.51 & 1.24 \\
22374-4550 & SKF 384 Aa,Ab &  & Acc9 & 0.2 & 0.78 & J & 0.18 & 0.82 & 1.01 & 5.72 & 5.63 \\
22526+5643 & WSI 10 & 112957 & Acc7 & 0.2 & 1.57 & G & 0.63\tablenotemark{\small b} & 0.91 & 0.534 & 4.27 & 8.00 \\
23464-2302 & TOK 375 & 117247 & Acc7 & 0.2 & 0.87 & H & 0.25 & 0.75 & 3.72 & 5.16 & 1.39 \\
23579-6805 & GKM9546 Aa,Ab & 118145 & Acc9 & 0.1 & 0.94 & G & 0.30 & 0.73 & 1.27 & 8.32 & 6.53 \\
\enddata
\tablenotetext{a}{Primary-mass source: G $=$ \gaia{} DR3 FLAME \citep{Fouesneau2023}; H $=$ \Hipparcos{} proper-motion-anomaly catalog \citep{Kervella2019}; J $=$ 2MASS $J$-band photometry.}
\tablenotetext{b}{No secondary photometry: $M_2 = q_{\rm typ}\,M_1$ and $f_{\rm pc}$ from the typical mass ratio of Section~\ref{sssec:m2_src} through Equation~\ref{eq:gamma}.}
\tablenotetext{c}{Boundary case, not one of the \nplaus plausible systems: $\aratio = 12.3$ places it just outside the window ($9.99$ without the photocenter correction). It is fitted in Section~\ref{ssec:orbitfit_results} and excluded from all population statistics.}
\tablecomments{The Discoverer column appends the component label for a resolved multiple (e.g., \texttt{Aa,Ab}). $\rho_1$ is the WDS first-epoch angular separation, $f_{\rm pc} = |1-\beta/B|$ the photocenter factor of Equation~\ref{eq:photocenter}, $\amax$ the photocenter-corrected face-on Newtonian upper bound (Section~\ref{ssec:amax}), $\ameas$ the total sky-plane \gaia{} NSS acceleration (Equation~\ref{eq:alpha_meas}), and $\aratio\equiv\ameas/\amax$.}
\end{deluxetable*}

\subsection{Orbit Fitting} \label{ssec:orbitfit}
For the systems in which the WDS cataloged companion can plausibly account for the measured acceleration, we attempt full Keplerian orbit fits with the \orvara{} package \citep{orvara}, run without epoch astrometry, so that the \Hipparcos{} and \gaia{} catalog quantities are evaluated from the model sky path over each mission window (Section~\ref{ssec:orvara_ext}).

The available absolute astrometry divides the sample into two groups according to whether the star has a \Hipparcos{} identifier, and we fit the two groups differently. For stars without a \Hipparcos{} entry, the DR3 acceleration is the only absolute-astrometric measurement available, and we fit it together with the resolved relative astrometry. We refer to this as Mode~A. For stars with a \Hipparcos{} entry we fit the DR3 acceleration jointly with the \Hipparcos--\gaia{} proper motion anomaly taken from the HGCA \citep{HGCA}, and refer to this as Mode~B.

The two modes differ not only in the number of data but in the time baselines of the constraints. The proper motion anomaly compares mean proper motions separated by roughly 25~yr and is therefore sensitive to orbits with periods of decades. The DR3 acceleration is a local measurement of the curvature of the sky path over the $\sim$34-month mission baseline, and its amplitude falls rapidly with increasing period. A Mode~A fit is consequently informative only for orbits short enough that the curvature is detectable within the \gaia{} baseline, and Mode~A solutions should be expected to favor short periods. This is a property of the data rather than of the sample, and we return to it in Section~\ref{ssec:orbitfit_results}.

One conversion of the absolute astrometry is required before fitting. \orvara{} models the reflex motion of the primary, whereas every absolute-astrometric quantity entering the likelihood refers to the photocenter of the unresolved pair: the \Hipparcos{} and \gaia{} proper motions, the \Hipparcos--\gaia{} proper motion, which is a difference of two photocenter positions, and the DR3 acceleration and jerk. As shown in Section~\ref{ssec:amax}, the displacement of the photocenter from the barycenter is that of the primary multiplied by $(1-\beta/B)$, so each of these quantities is the primary's multiplied by the factor $\fpc$ of Equation~\ref{eq:photocenter}. The identity holds at every instant and therefore for every time derivative and every finite difference of position. We accordingly divide the five absolute-astrometric quantities and their uncertainties by $\fpc$, taking the value tabulated in Table~\ref{tab:plausible}, and leave the parallax, the correlation coefficients, and the relative astrometry unchanged, the last because a secondary-minus-primary measurement carries no photocenter. Dividing the absolute proper motions is permissible because \orvara{} treats the barycentric proper motion as a free parameter and marginalizes over it analytically. After the division each proper motion equals the reflex motion of the primary plus a constant, the barycentric motion divided by $\fpc$, and that constant is absorbed by the free parameter. Dividing the uncertainties by the same factor leaves the signal-to-noise ratio of every quantity unchanged, so the relative weight of each measurement in the likelihood is as it was. The accuracy of the conversion is limited by the magnitude difference, which enters $\fpc$ through both of its ingredients: the photocenter position $\beta$ follows from $\Delta m$ directly, and the barycenter position $B$ follows from the mass ratio $q$, which is itself derived from $\Delta m$ through the mass--luminosity relation ($\beta$ and $B$ are defined in Section~\ref{ssec:amax}). An error of $0.3$~mag in $\Delta m$ therefore changes $\fpc$ by $17\%$ at $\Delta m = 1.5$~mag and by $30$--$45\%$ at $\Delta m \le 1$~mag. This is the dominant systematic on the companion masses of the closest pairs, and we quantify it in Section~\ref{ssec:masses} by refitting those systems with $\fpc$ recomputed at $\Delta m + 0.3$~mag.

Mode~A requires an additional prior. With two to five resolved epochs and the DR3 coefficients as the only absolute-astrometric data, the likelihood is flat in the companion mass over orders of magnitude, and an unbounded chain wanders to the edge of the prior. In our first Mode~A fits the walkers reached $M_2 = 30$--$1000\,\Msun$, the posterior medians were set by the logarithmic prior rather than by the data, and one chain, that of 19169$-$4348, never became stationary. We therefore place a Gaussian prior on the companion mass in Mode~A, centered on the photometric $M_2$ of Table~\ref{tab:plausible} with a width of half that value. The Mode~A companion masses are consequently not measurements, and we do not tabulate the Mode~A orbits. The Mode~A fits test only whether the DR3 coefficients can be reproduced by a Keplerian orbit with a physically admissible companion. Mode~B carries no such prior. In both modes the primary mass carries a Gaussian prior centered on the estimate of Section~\ref{ssec:m1} with a width of 10\% of that value.

We sample with \orvara{}'s parallel-tempered MCMC as in \citet{156_systems}, here with 20 temperature levels and 100 walkers per level, $2\times10^{5}$ steps per walker, and thinning by a factor of 200. Inference uses the coldest chain, of which we discard the initial half as burn-in. Convergence is checked individually for each parameter: for these 100-walker, 500-step (thinned, post-burn-in) chains the integrated autocorrelation time is $\le1.4$ thinned steps for every Keplerian parameter in all \Nfit fits, so each posterior rests on $\gtrsim$$3.5\times10^{4}$ quasi-independent samples, and the log-likelihood and the companion-mass median are stationary across the retained chain. We additionally inspect \orvara{}'s diagnostic output for every system to confirm that no systematic residual is concealed by the summary statistics. The differences in fit quality between systems are therefore differences in how well the model reproduces the data, not artifacts of the sampling.

Two error conventions enter the likelihood. The WDS measures carry heterogeneous and often nominal uncertainties, so for each system with at least seven resolved epochs we scale the separation and the position-angle uncertainties separately, by $\sqrt{\chi^2_\rho/N_{\rm rel}}$ and $\sqrt{\chi^2_{\rm PA}/N_{\rm rel}}$ evaluated at the maximum-likelihood step of the preceding fit, and re-fit. We never reduce the cataloged uncertainties, so a component whose nominal errors are already conservative retains them (the sub-unity entries in Table~\ref{tab:chisq}), and we leave the systems with fewer than seven epochs unscaled, since two to four epochs do not determine a scaling factor. The scaling is applied once, to six of the nine Mode~B systems, with factors between $1.07$ and $3.04$. The DR3 acceleration and jerk uncertainties are the marginal $1\sigma$ values of the NSS covariance, inflated by a factor of two following the calibration of \citet{156_systems}. Two individual measures, of 11585$-$7754 at 2016.05 and of 23464$-$2302 at 2015.91, have position angles $180\degr$ from the adjacent measures and from the published orbit ephemerides in the WDS files, and we rotate these two by $180\degr$ for the fits. The direction test of Section~\ref{ssec:dirtest} uses every measure as recorded.

\subsection{Direction Test} \label{ssec:dirtest}

If the cataloged companion produces the \gaia{} acceleration, the acceleration must point toward the companion, since the primary accelerates toward the companion and the photocenter, which lies on the primary's side of the barycenter whenever $B > \beta$, accelerates in the same direction (Section~\ref{ssec:amax}). We therefore compare $\mathrm{PA}_{\rm acc}$ (Section~\ref{ssec:ratio}) with the companion position angle. The test uses a direction only. No separation or orbit enters it, so it applies to every system with a position angle rather than only to those with converged fits, and it is free of the separation systematic that dominates $\aratio$.

The companion position angle is taken from the WDS measurement files (Section~\ref{ssec:wds}), whose individual measurement times can be compared to the \gaia{} epoch. For each pair we adopt the position angle measurement closest to J2016.0 from the block of the cross-matched component pair. Propagating the position angle to J2016.0 at the rate implied by the two nearest measures changes nothing of substance (Section~\ref{ssec:direction}), so we use the measure as recorded.

The condition $B > \beta$ of Section~\ref{ssec:amax} holds for any main-sequence or degenerate companion fainter than its primary, since $L \propto M^{\gamma}$ with $\gamma > 1$ gives $\phi = q^{\gamma} < q$, so the expected offset is zero. We nevertheless fold the offset onto the axis, $|\Delta\theta| \equiv |\mathrm{PA}_{\rm acc} - \mathrm{PA}_{\rm WDS}| \in [0\degr, 90\degr]$, to allow for the $180\degr$ ambiguity of the cataloged position angles. Most of these close pairs are measured by speckle interferometry, which recovers the autocorrelation of the pair and therefore its axis but not the side on which the companion lies. The quadrant is assigned afterward, and for stars within a magnitude of each other it is often a guess. The fold accommodates this property of the catalog, not a physical freedom. An acceleration that points away from the cataloged companion is either a reversed quadrant or a pair in which the \gaia{}-tracked light lies on the companion's side of the barycenter, and the orbit fits, which apply no fold, distinguish the two for the fitted systems (Section~\ref{ssec:direction}). Under the fold a random acceleration falls within $45\degr$ of the axis half of the time, so the uniform expectation is $50\%$. We quantify the departure from uniformity with two tests. The first is the binomial probability of obtaining at least the observed number of systems within $45\degr$ when each has probability $50\%$. The second is a Kolmogorov--Smirnov test of the full distribution of $|\Delta\theta|$ against the uniform distribution on $[0\degr, 90\degr]$, which does not depend on the $45\degr$ threshold (Section~\ref{ssec:direction}).

\section{Results}\label{sec:results}
\begin{figure*}
    \centering
    {\fig{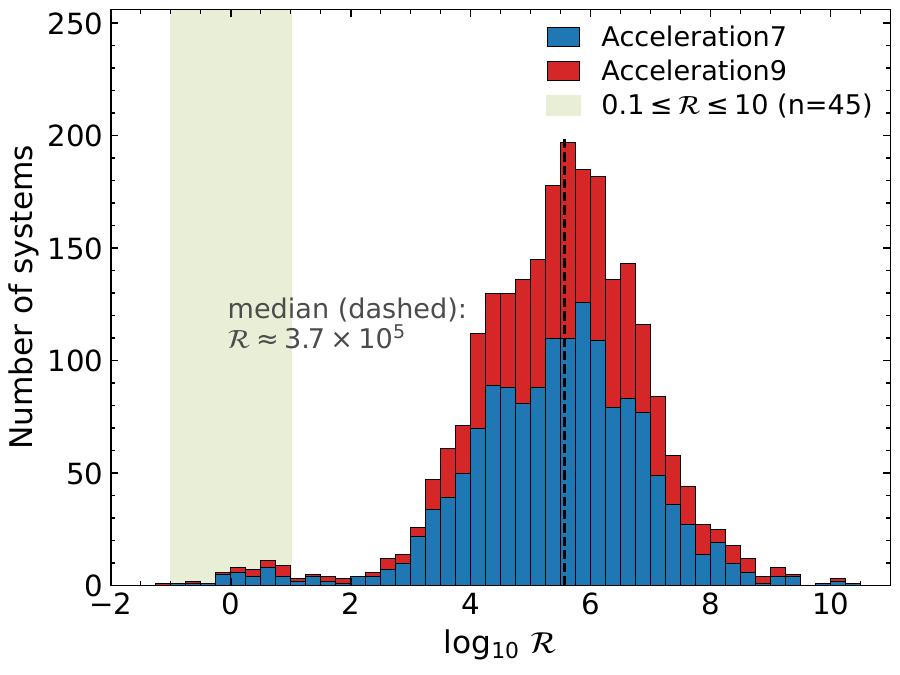}{0.48\textwidth}{(a)}
    \fig{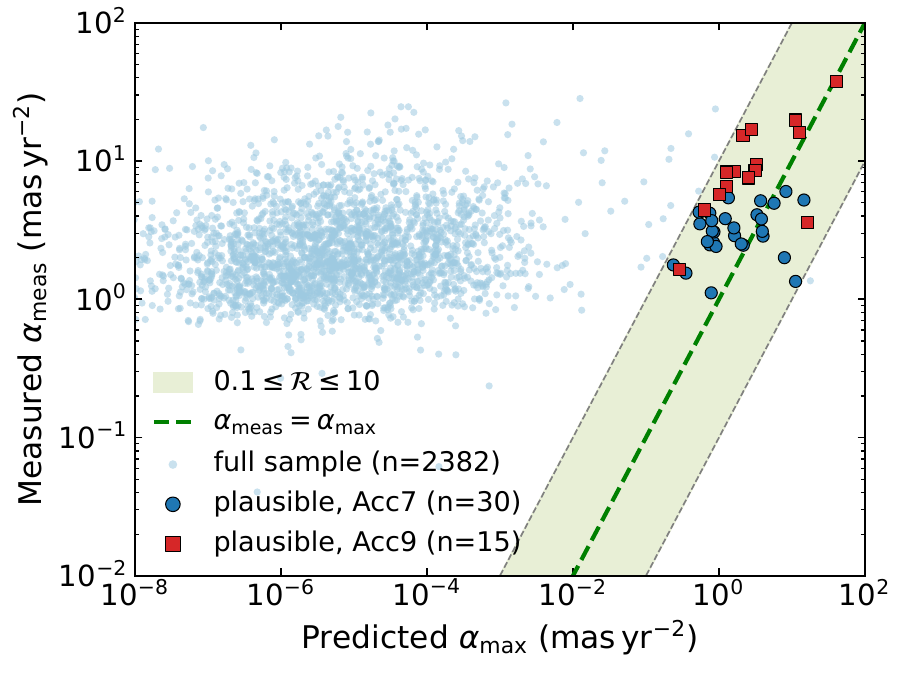}{0.48\textwidth}{(b)}}
    \caption{\label{fig:fullsample}Full-sample acceleration-ratio diagnostics for the \nfinite cross-matched systems with a finite ratio (the \nnoratio pairs without a finite ratio, Section~\ref{ssec:ratio}, are excluded). \textit{(a)}: Distribution of the acceleration ratio $\aratio = \ameas/\amax$, split and colored by \gaia{} NSS solution type (\texttt{Acceleration7} in blue and \texttt{Acceleration9} in red, stacked). The dashed line marks the median, $\aratio \approx \medRfull\times10^{5}$. Most systems have $\aratio \gg 1$, consistent with these signals being driven by closer, unresolved perturbers. \textit{(b)}: Measured astrometric acceleration $\ameas$ versus the predicted maximum acceleration $\amax$, the photocenter-corrected Newtonian bound of Equation~\ref{eq:photocenter}. Light points show the full finite sample. The dashed line is the one-to-one relation $\ameas=\amax$. In both panels the shaded band marks the physically plausible range $0.1\le\aratio\le10$, which isolates the \nplaus wide binaries (\texttt{Acceleration7}, blue circles, $n=\nplausa$, and \texttt{Acceleration9}, red squares, $n=\nplausb$). The \nhigher systems above the band have $\aratio>10$. Two systems lie just outside it, 18267$-$3024 below at $\aratio = 0.08$ and 16206$+$4535 above at $\aratio = \Rboundary$ (Section~\ref{ssec:ratio}).}
\end{figure*}

The acceleration-ratio selection of Section~\ref{ssec:ratio} reduces the \nuniq cross-matched pairs to \nplaus plausible wide binaries in which the cataloged companion can account for the measured acceleration (median $\aratio = \medR$). These are close pairs: the median projected separation is $\medsepAU$~AU (range $\minsepAU$--$\maxsepAU$~AU) at a median distance of $\meddpc$~pc. Figure~\ref{fig:fullsample} shows the full-sample distribution of the acceleration ratio and the measured acceleration against the Newtonian bound, and Table~\ref{tab:plausible} lists the plausible systems. This section presents the orbit fits (Section~\ref{ssec:orbitfit_results}), the recovery of the DR3 coefficients (Section~\ref{ssec:coeffs}) and the one joint fit that fails it (Section~\ref{ssec:14382}), the comparison of dynamical and photometric companion masses (Section~\ref{ssec:masses}), the direction test (Section~\ref{ssec:direction}), and the changes made to \orvara{} and \texttt{htof} (Section~\ref{ssec:orvara_ext}).

\subsection{Orbit Fit Results} \label{ssec:orbitfit_results}

The \nplauship plausible systems with \Hipparcos{} counterparts are fit in Mode~B and the remaining \nplausnohip in Mode~A (Section~\ref{ssec:orbitfit}). Converged fits are obtained for \nfit systems, nine in Mode~B and eight in Mode~A: sixteen plausible systems and the boundary case 16206$+$4535 of Section~\ref{ssec:ratio} (Table~\ref{tab:orbits}, note~a). The remaining 29 plausible systems remain in the plausible sample and in the direction test of Section~\ref{ssec:direction} but are set aside from the orbit fitting, for the following reasons:

\begin{itemize}
    \item \textit{Cataloged pair not the source of the signal} (4). In 12104$-$4352 (HIP 59341) and 16238$-$0258 (HIP 80315) an inner companion constrained by radial velocities and absolute astrometry produces the acceleration, in 23579$-$6805 (HIP 118145) the perturber is unresolved, and 20467$+$1846, whose single WDS entry no later measurement confirms, is most likely an unresolved single star rather than a pair. These lie outside the test posed here.
    \item \textit{Acceleration not constant across the DR3 baseline} (2). The J2016.0 value does not characterize the orbit, a property of the orbit rather than a failure of the \gaia{} astrometry. 11253$-$8457 (HIP 55746) has a published $P = 8.98 \pm 0.64$~yr, $e = 0.21$ orbit \citep{Tok2022g} and a separation falling from $0\farcs061$ at 2016.0 to $0\farcs027$ at 2017.4, and 02136$-$0849 (HIP 10365) has an \texttt{Acceleration9} \texttt{significance} of 92.2, exceeded by only 111 of the $91{,}268$ such solutions in DR3, and a separation falling from $0\farcs056$ in 2012.7 to $0\farcs016$ in 2023.9.
    \item \textit{Too sparse to fit} (\nsparse). Confirmed pairs with too few resolved measurements to constrain a seven-parameter Keplerian orbit. Two have no usable measurement, only occultation timings (08423$+$1322 and 20335$-$1333). Fourteen have a single usable measurement (04248$+$1552, 04433$+$1704, 05369$-$4758, 06119$-$3529, 09293$-$4432, 10095$-$3251, 11181$+$1533, 13097$-$2212, 16334$-$2015, 18060$-$4445, 19156$-$0830, 22084$-$3259, 22246$+$3926 and 22526$+$5643), five of them last measured in 2005 or earlier. Seven have two or three usable measurements (04476$+$1816, 12146$-$1517, 14115$-$1204, 14460$+$4633, 15210$-$1533, 20192$-$2610 and 22006$-$1410), either confined to 2022--2024 or, for 15210$-$1533, an occultation timing and a speckle measure a decade apart. Fits attempted for the two-epoch systems 04476$+$1816 and 22006$-$1410 did not converge.
\end{itemize}

Sparseness of the WDS measures is the dominant attrition channel and the reason the orbit-fitting sample is less than half the plausible sample. It is a limitation of the resolved astrometry, not of the \gaia{} accelerations, and for these systems the direction test of Section~\ref{ssec:direction}, which requires only a position angle, is the only per-system check available.

\setlength{\tabcolsep}{4pt}
\begin{deluxetable*}{lcccccccccc}
\tabletypesize{\scriptsize}
\tablecaption{Relative-astrometry coverage and posterior orbital parameters for the nine Mode~B fits.\label{tab:orbits}}
\tablehead{
\colhead{WDS} &
\colhead{$N_{\rm rel}$} & \colhead{$\Delta t/P$} & \colhead{$\Delta m$} &
\colhead{$M_1$} & \colhead{$M_2$} & \colhead{$M_{\rm dyn}/M_{\rm phot}$} & \colhead{$a$} &
\colhead{$i$} & \colhead{$P$} & \colhead{$e$} \\[-0.6em]
\colhead{} &
\colhead{} & \colhead{} & \colhead{(mag)} &
\colhead{($\Msun$)} & \colhead{($\Msun$)} & \colhead{} & \colhead{(AU)} &
\colhead{(deg)} & \colhead{(yr)} & \colhead{}
}
\startdata
03309$-$6200 & 15 & 3.54 & 1.5 & ${1.35}_{-0.11}^{+0.11}$ & ${1.034}_{-0.060}^{+0.061}$ & 1.13 & ${3.118}_{-0.070}^{+0.068}$ & ${81.61}_{-0.40}^{+0.39}$ & ${3.565}_{-0.023}^{+0.023}$ & ${0.527}_{-0.028}^{+0.028}$ \\
16077$-$2125\tablenotemark{\small c} & 8 & 1.28 & 1.0 & ${1.95}_{-0.19}^{+0.20}$ & ${2.24}_{-0.50}^{+0.80}$ & 1.41 & ${8.20}_{-0.34}^{+0.50}$ & ${82.36}_{-0.73}^{+0.74}$ & ${11.44}_{-0.20}^{+0.24}$ & ${0.399}_{-0.047}^{+0.098}$ \\
16206$+$4535\tablenotemark{\small a} & 10 & 0.64 & 2.9 & ${1.225}_{-0.093}^{+0.095}$ & ${0.896}_{-0.037}^{+0.037}$ & 1.39 & ${12.41}_{-0.24}^{+0.24}$ & ${31.3}_{-2.5}^{+2.1}$ & ${30.00}_{-0.34}^{+0.38}$ & ${0.6860}_{-0.0075}^{+0.0073}$ \\
05231$-$3142\tablenotemark{\small c} & 3 & 0.18 & 0.9 & ${0.909}_{-0.093}^{+0.093}$ & ${1.53}_{-0.21}^{+0.26}$ & 2.03 & ${6.15}_{-0.24}^{+0.26}$ & ${106.3}_{-1.4}^{+1.5}$ & ${9.761}_{-0.078}^{+0.082}$ & ${0.590}_{-0.020}^{+0.020}$ \\
11585$-$7754\tablenotemark{\small c} & 7 & 1.80 & 0.7 & ${0.986}_{-0.092}^{+0.091}$ & ${1.675}_{-0.082}^{+0.082}$ & 2.03 & ${8.39}_{-0.26}^{+0.29}$ & ${73.47}_{-0.85}^{+0.83}$ & ${14.89}_{-0.42}^{+0.51}$ & ${0.615}_{-0.019}^{+0.019}$ \\
14382$+$1402\tablenotemark{\small d} & 15 & 1.06 & 1.5 & ${1.118}_{-0.082}^{+0.084}$ & ${0.973}_{-0.047}^{+0.049}$ & 1.22 & ${5.41}_{-0.11}^{+0.11}$ & ${116.10}_{-0.93}^{+0.96}$ & ${8.710}_{-0.027}^{+0.027}$ & ${0.424}_{-0.011}^{+0.012}$ \\
23464$-$2302\tablenotemark{\small e} & 14 & 1.06 & 3.1 & ${0.763}_{-0.065}^{+0.066}$ & ${0.516}_{-0.031}^{+0.032}$ & 2.03 & ${5.81}_{-0.14}^{+0.13}$ & ${147.4}_{-3.1}^{+3.5}$ & ${12.38}_{-0.11}^{+0.10}$ & ${0.518}_{-0.016}^{+0.016}$ \\
16268$-$4936 & 4 & 0.14 & 3.8 & ${0.463}_{-0.043}^{+0.045}$ & ${0.130}_{-0.011}^{+0.011}$ & 0.94 & ${7.60}_{-0.25}^{+0.25}$ & ${115.5}_{-2.5}^{+2.2}$ & ${27.24}_{-0.71}^{+0.70}$ & ${0.180}_{-0.041}^{+0.052}$ \\
08358$+$0637\tablenotemark{\small b} & 6 & 0.18 & \nodata & ${0.828}_{-0.088}^{+0.089}$ & ${0.0683}_{-0.0087}^{+0.0124}$ & \nodata & ${6.15}_{-0.34}^{+3.03}$ & ${59.3}_{-3.2}^{+6.4}$ & ${16.2}_{-1.5}^{+13.3}$ & ${0.43}_{-0.13}^{+0.12}$ \\
\enddata
\tablenotetext{a}{Boundary case (Section~\ref{ssec:ratio}): 16206$+$4535 has $\aratio = 12.3$, just outside the plausible window.}
\tablenotetext{b}{No WDS secondary magnitude: $\fpc = 0.74$ from the assumed mass ratio of Section~\ref{sssec:m2_src}. The companion is HD~72946~B (Section~\ref{ssec:masses}).}
\tablenotetext{c}{$\Delta m \le 1$~mag: at the cataloged $\Delta m$ the photocenter conversion returns $M_2 > M_1$, which a fainter main-sequence companion cannot have, so the mass is bounded by the photometry rather than measured. Recomputing $\fpc$ at $\Delta m + 0.3$~mag lowers $M_2$ by 26--41\% (Section~\ref{ssec:masses}).}
\tablenotetext{d}{Reproduces the DR3 acceleration but not the jerk, and leaves systematic position-angle residuals (Table~\ref{tab:chisq}, Section~\ref{ssec:14382}).}
\tablenotetext{e}{Twice the photometric mass. The dynamical $0.52\,\Msun$ lies on the $\gamma = 4$ branch of Equation~\ref{eq:gamma} whereas the photometric solution fell on the $\gamma = 2.3$ branch, and an unresolved pair in place of a single companion would raise the mass at fixed $\Delta m$ by a further factor of up to $1.5$--$1.7$ (Section~\ref{ssec:masses}).}
\tablecomments{Posterior medians with 16th/84th-percentile uncertainties. HIP identifiers and solution types are in Tables~\ref{tab:plausible} and \ref{tab:chisq}. $N_{\rm rel}$: WDS relative-astrometry epochs. $\Delta t/P$: fraction of the median-period orbit they span. $\Delta m$: WDS magnitude difference. $M_{\rm dyn}/M_{\rm phot}$: fitted companion mass over the photometric estimate of Table~\ref{tab:plausible}.}
\end{deluxetable*}
%% chi^2 by data type at the maximum-likelihood step. Rows generated 2026-09-26 by
%% orbit_fit_prep/postprocess_chain.py from the same chains as table_orbits.tex (08358+0637 added 2026-10-03);
%% chi2_rho/N_rel and chi2_PA/N_rel separately; nu_G = 4/6 (Mode B), 2/4 (Mode A).
\setlength{\tabcolsep}{3.5pt}
\begin{deluxetable*}{llcccccccccccc}
\tabletypesize{\scriptsize}
\tablecaption{Contribution of each data type to $\chi^2$ at the maximum-likelihood step of every converged fit.\label{tab:chisq}}
\tablehead{
\colhead{WDS} & \colhead{HIP} & \colhead{Mode} & \colhead{Sol.} & \colhead{$N_{\rm rel}$} &
\colhead{$\chi^2_\rho$} & \colhead{$\chi^2_{\rm PA}$} & \colhead{$\chi^2_\rho/N_{\rm rel}$} & \colhead{$\chi^2_{\rm PA}/N_{\rm rel}$} &
\colhead{$\chi^2_{\rm H}$} & \colhead{$\chi^2_{\rm HG}$} & \colhead{$\chi^2_{\rm G}$} & \colhead{$\nu_{\rm G}$} & \colhead{$\chi^2_{\rm G}/\nu_{\rm G}$}
}
\startdata
03309$-$6200 & 16370 & B & Acc9 & 15 & 3.9 & 17.1 & 0.26 & 1.14 & 12.1 & 0.02 & 0.6 & 6 & 0.11 \\
16077$-$2125 & 79020 & B & Acc7 & 8 & 8.6 & 9.3 & 1.08 & 1.16 & 2.1 & 0.11 & 0.4 & 4 & 0.09 \\
16206$+$4535 & 80074 & B & Acc9 & 10 & 4.9 & 10.4 & 0.49 & 1.04 & 7.2 & 0.02 & 0.8 & 6 & 0.13 \\
05231$-$3142 & 25176 & B & Acc9 & 3 & 1.1 & 1.3 & 0.35 & 0.42 & 3.7 & 0.08 & 0.1 & 6 & 0.02 \\
11585$-$7754 & 58400 & B & Acc7 & 7 & 6.8 & 7.5 & 0.97 & 1.07 & 1.9 & 0.13 & 0.7 & 4 & 0.18 \\
14382$+$1402 & 71572 & B & Acc9 & 15 & 12.8 & 34.0 & 0.85 & 2.26 & 7.5 & 0.35 & 98.6 & 6 & 16.44 \\
23464$-$2302 & 117247 & B & Acc7 & 14 & 16.7 & 11.1 & 1.19 & 0.79 & 1.1 & 0.14 & 15.3 & 4 & 3.82 \\
16268$-$4936 & 80559 & B & Acc7 & 4 & 7.9 & 3.6 & 1.97 & 0.90 & 0.5 & 0.02 & 0.3 & 4 & 0.09 \\
08358$+$0637 & 42173 & B & Acc7 & 6 & 0.2 & 0.0 & 0.03 & 0.00 & 4.4 & 0.24 & 0.3 & 4 & 0.07 \\
\hline
19169$-$4348 & \nodata & A & Acc9 & 3 & 27.1 & 1.2 & 9.05 & 0.38 & \nodata & \nodata & 6.6 & 4 & 1.64 \\
16387$-$3605 & \nodata & A & Acc7 & 2 & 0.0 & 0.0 & 0.01 & 0.02 & \nodata & \nodata & 0.1 & 2 & 0.03 \\
08588$-$5303 & \nodata & A & Acc7 & 2 & 0.1 & 0.2 & 0.04 & 0.08 & \nodata & \nodata & 0.1 & 2 & 0.03 \\
16252$-$3048 & \nodata & A & Acc9 & 2 & 0.0 & 0.3 & 0.01 & 0.13 & \nodata & \nodata & 2.3 & 4 & 0.58 \\
18027$-$5714 & \nodata & A & Acc7 & 2 & 0.8 & 0.0 & 0.41 & 0.00 & \nodata & \nodata & 0.1 & 2 & 0.07 \\
22374$-$4550 & \nodata & A & Acc9 & 5 & 2.0 & 0.7 & 0.40 & 0.14 & \nodata & \nodata & 1.4 & 4 & 0.35 \\
04407$-$2812 & \nodata & A & Acc9 & 4 & 1.3 & 0.5 & 0.32 & 0.13 & \nodata & \nodata & 0.1 & 4 & 0.03 \\
00350$-$6342 & \nodata & A & Acc7 & 2 & 0.1 & 0.8 & 0.06 & 0.39 & \nodata & \nodata & 0.0 & 2 & 0.02 \\
\enddata
\tablecomments{$\chi^2_\rho$ and $\chi^2_{\rm PA}$ are summed over the $N_{\rm rel}$ resolved epochs after the uncertainty scaling of Section~\ref{ssec:orbitfit}. $\chi^2_{\rm G}$ is evaluated against the photocenter-converted DR3 coefficients.}
\end{deluxetable*}

\begin{figure*}
\centering
\includegraphics[width=0.92\textwidth]{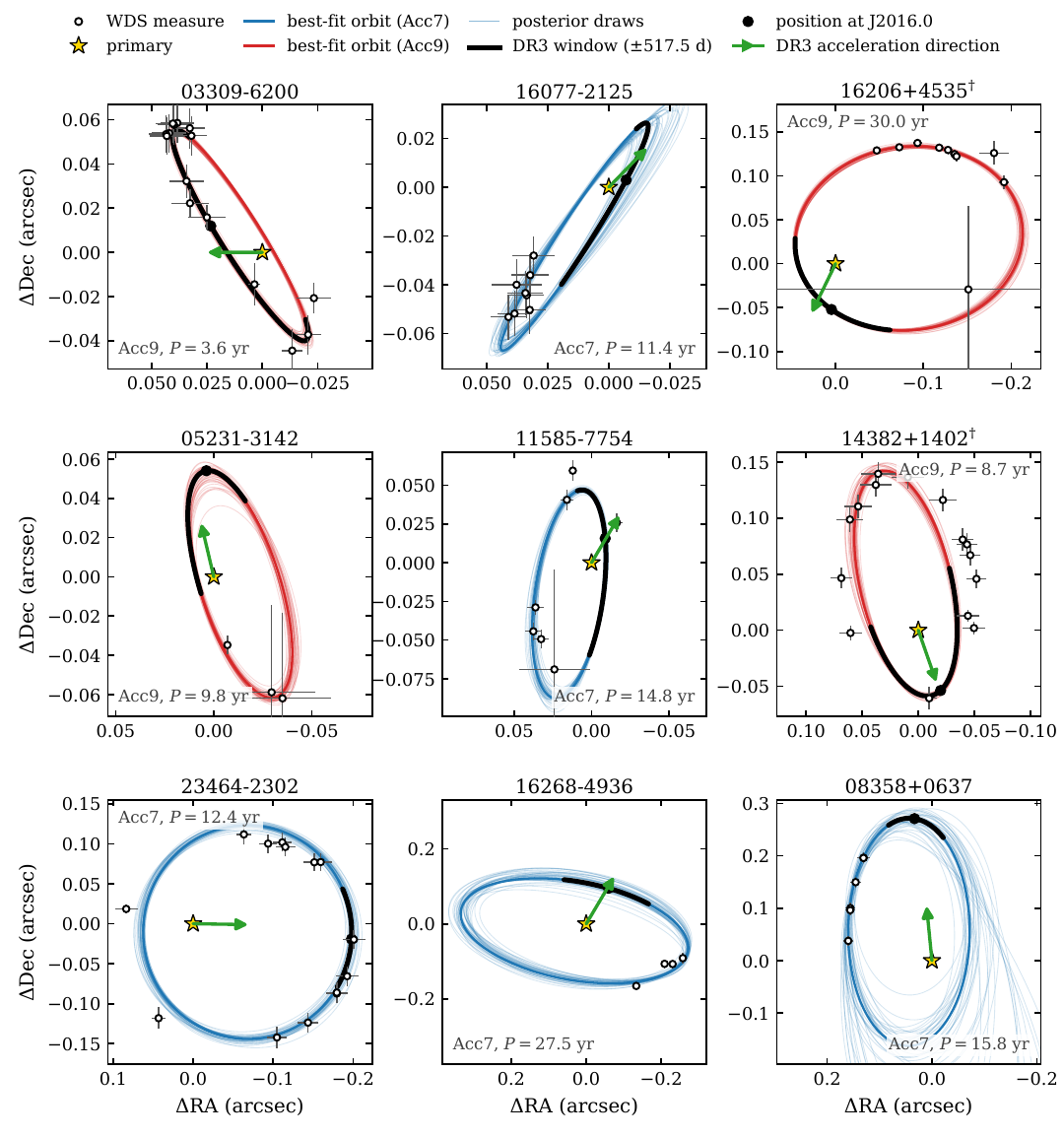}
\caption{Relative orbits of the nine Mode~B systems. Points are the WDS measurements of the secondary relative to the primary with the uncertainties adopted in the fit. The solid curve is the orbit at the maximum-likelihood step of the final chain, and the thin curves are 40 orbits drawn at random from the post-burn-in posterior. The thick black arc is the DR3 window, J2016.0 $\pm$ 517.5~d, with the position at J2016.0 marked. The green arrow from the primary gives the direction of the cataloged DR3 acceleration, with arbitrary length. North is up and east is left. Daggers mark 16206$+$4535, the boundary case of Table~\ref{tab:orbits}, and 14382$+$1402, whose fit does not reproduce the DR3 jerk. The two posterior modes of 08358$+$0637 ($P = 16$ and $32$~yr, Section~\ref{ssec:masses}) are both visible in its draws, and the large cross in the 16206$+$4535 panel is the 1991.25 \Hipparcos{} discovery measure with its adopted $0\farcs5$ uncertainty.}
\label{fig:orbits}
\end{figure*}

Table~\ref{tab:orbits} lists the nine Mode~B fits and Figure~\ref{fig:orbits} shows their orbits against the WDS measures. Their periods follow the division of Section~\ref{sec:intro}: 11--27~yr for the five \texttt{Acceleration7} systems and 3.6--9.8~yr for three of the four \texttt{Acceleration9} systems, the fourth, 16206$+$4535, being a 30-yr orbit with $e = 0.69$ that passes periastron inside the DR3 window.

The Mode~A fits are not tabulated: with two to five epochs covering 3--31\% of the fitted period, their posteriors are set by the companion-mass prior of Section~\ref{ssec:orbitfit} and by the geometry of the arc, and the discussion of masses and elements below is restricted to Mode~B. The leverage of the Mode~B fits on the primary mass is modest. The posteriors on $M_1$ have 0.69--1.02 of the width of the prior, and only 23464$-$2302 ($-1.2\sigma$) and 16268$-$4936 ($-2.7\sigma$) are displaced from it by more than $1\sigma$, where the relative orbit and the parallax, through Kepler's third law, pull against the photometric estimate.

\subsection{Recovery of the DR3 Coefficients} \label{ssec:coeffs}

Table~\ref{tab:chisq} lists the contribution of each data type to $\chi^2$ at the maximum-likelihood step of every fit. The \gaia{} term $\chi^2_{\rm G}$ comprises the DR3 proper motion, acceleration and, for \texttt{Acceleration9} sources, the jerk, with $\nu_{\rm G} = 4$ or $6$ in Mode~B. In Mode~A the barycentric proper motion is a free parameter constrained only by the DR3 proper motion, which it reproduces exactly, so $\nu_{\rm G} = 2$ or $4$. For the six Mode~B systems with $N_{\rm rel} \ge 7$, the uncertainty scaling of Section~\ref{ssec:orbitfit} brings $\chi^2_\rho/N_{\rm rel}$ and $\chi^2_{\rm PA}/N_{\rm rel}$ to about unity by construction, so these entries measure the scaling rather than the fit. The one exception is the position-angle term of 14382$+$1402, which remains at $2.26$ after the scaling.

\begin{figure*}
\centering
\includegraphics[width=0.48\textwidth]{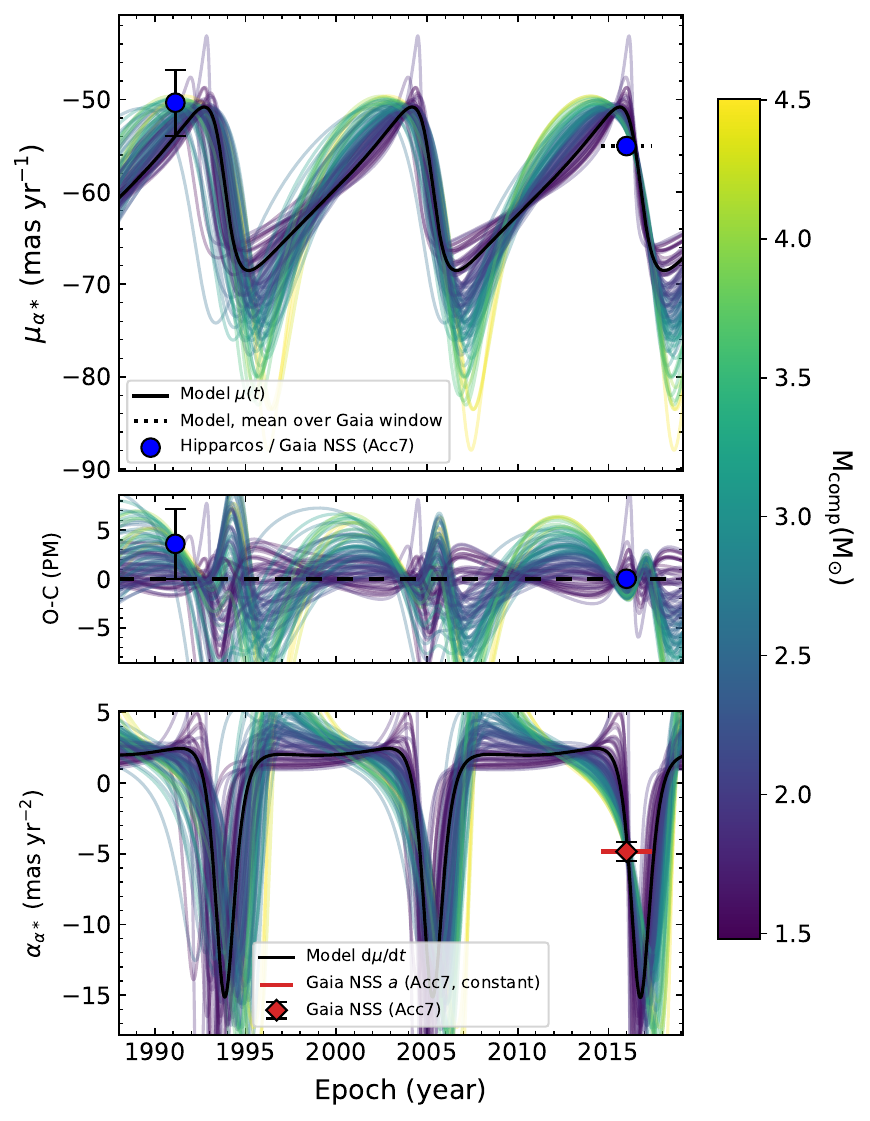}\hfill
\includegraphics[width=0.48\textwidth]{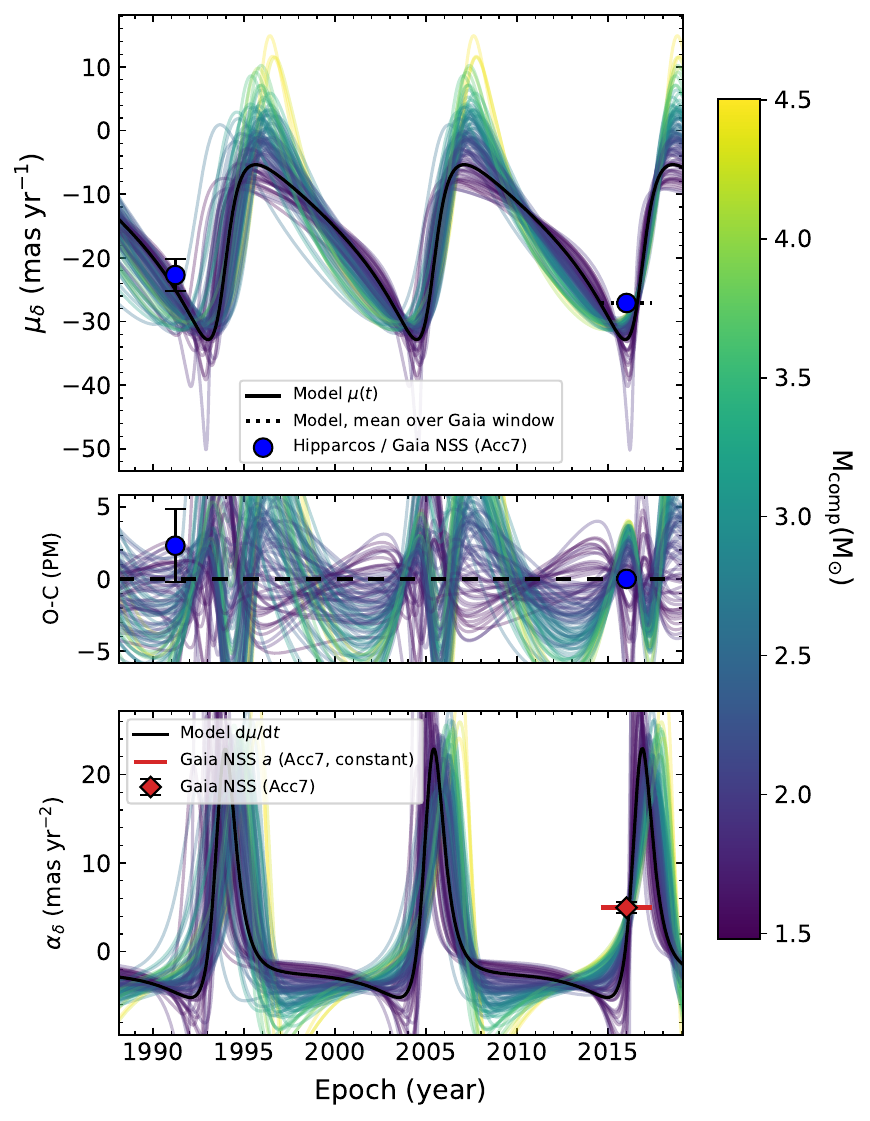}
\caption{The \orvara{} proper-motion diagnostic of Section~\ref{ssec:orvara_ext} for 16077$-$2125 (\texttt{Acceleration7}, Mode~B), in right ascension (left) and declination (right). In each column the upper panel is the model proper motion against epoch, with the maximum-likelihood orbit in black and posterior draws colored by companion mass, the \Hipparcos{} and \gaia{} measurements as points, and the mean of the model over the DR3 window as a dotted segment, with the residuals in the panel below. The lower panel is the time derivative of the curve above, the model acceleration, with the cataloged constant and its $1\sigma$ band drawn in red over the window $2016.0 \pm 517.5$~d. The fit reproduces the cataloged acceleration ($\chi^2_{\rm G}/\nu_{\rm G} = 0.09$, Table~\ref{tab:chisq}). The relative orbit, the separation and position-angle residuals, and the corner plot of the posterior of this fit are in Appendix~\ref{app:fits}.}
\label{fig:set16077}
\end{figure*}

\begin{figure*}
\centering
\includegraphics[width=0.48\textwidth]{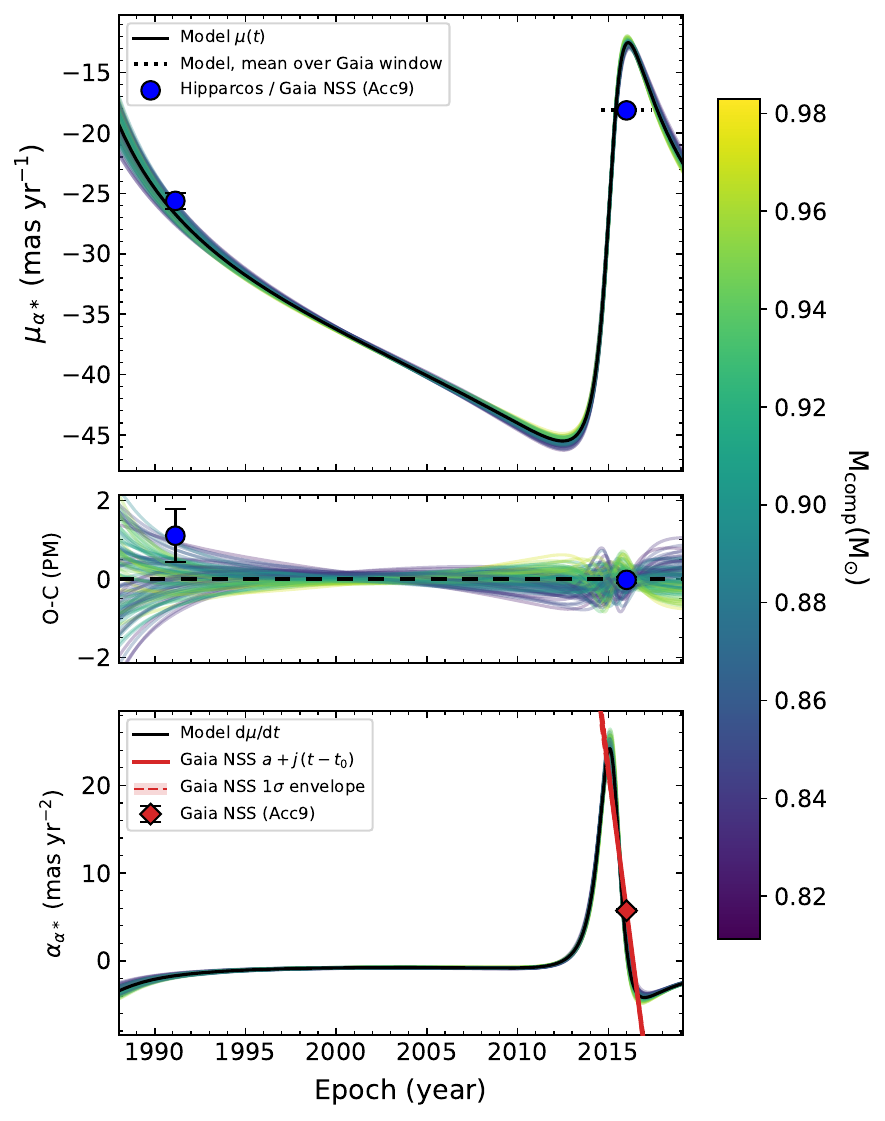}\hfill
\includegraphics[width=0.48\textwidth]{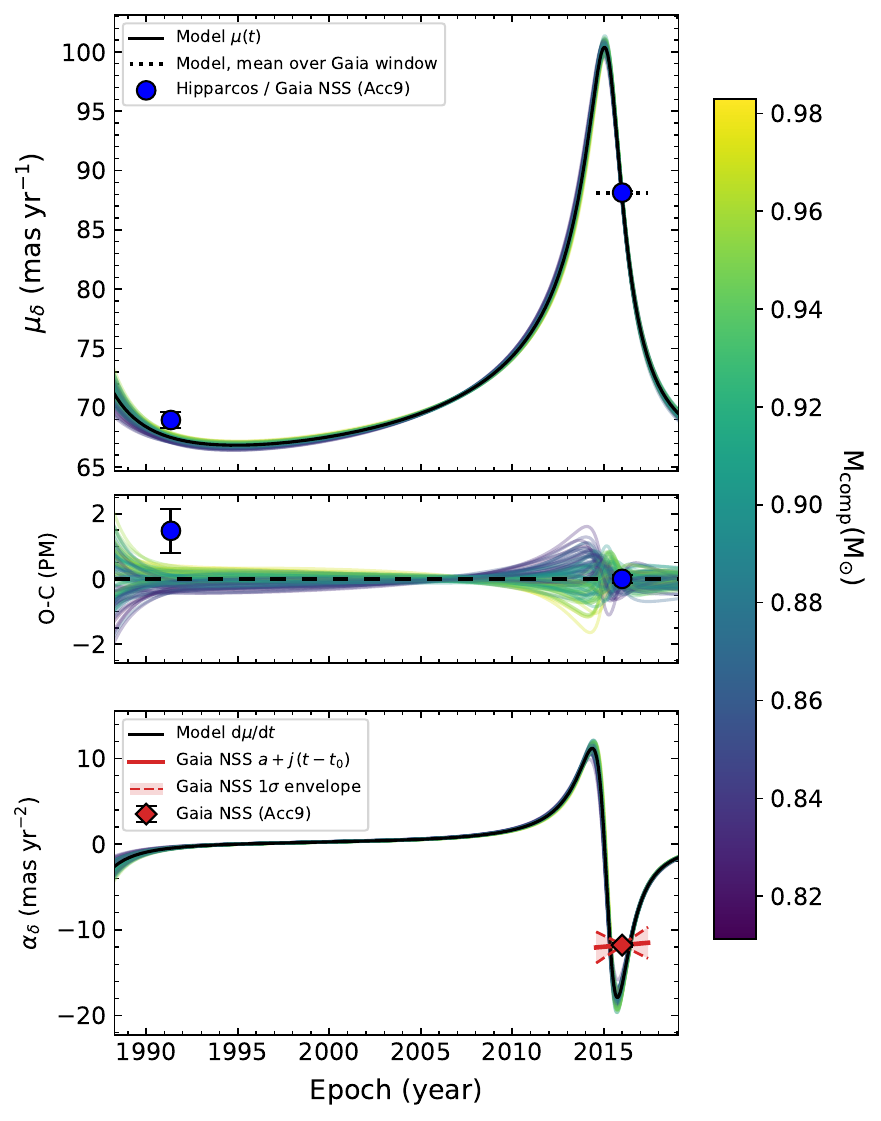}
\caption{As Figure~\ref{fig:set16077}, for 16206$+$4535 (\texttt{Acceleration9}, Mode~B), the boundary case of Table~\ref{tab:orbits}, note~a. For a nine-parameter solution the DR3 measurement in the acceleration panels is the segment $a + j\,(t - \mathrm{J2016.0})$ with the envelope from the two uncertainties, so the cataloged jerk is the slope of the segment. The fit reproduces the acceleration and the jerk together ($\chi^2_{\rm G}/\nu_{\rm G} = 0.13$), the segment in right ascension following the model curve through the periastron passage inside the window. In declination the jerk is consistent with zero and the model curve dips below the flat segment inside the window, because the curve is the instantaneous acceleration whereas the likelihood compares finite differences across the window (Section~\ref{ssec:orvara_ext}). The relative orbit, the separation and position-angle residuals, and the corner plot of the posterior of this fit are in Appendix~\ref{app:fits}.}
\label{fig:set16206}
\end{figure*}

Eight of the nine Mode~B fits reproduce all three data sets, seven of them with $\chi^2_{\rm G}/\nu_{\rm G} \le 0.2$ and 23464$-$2302 with $3.8$. The ninth, 14382$+$1402, reproduces the \Hipparcos{} and \Hipparcos--\gaia{} proper motions ($\chi^2_{\rm H} + \chi^2_{\rm HG} = 7.9$ for four degrees of freedom), the resolved orbit, and the DR3 acceleration, but not the jerk ($\chi^2_{\rm G}/\nu_{\rm G} = 16.4$, of which the two jerk components contribute $79$ of the $98.6$), and it leaves systematic residuals in the resolved position angles. Among the Mode~A fits, the four \texttt{Acceleration7} fits carry no test of the coefficients: two resolved epochs and the two acceleration components are six data for seven orbital parameters, so an orbit reproducing every input exists whether or not the acceleration is correct ($\chi^2_{\rm G}/\nu_{\rm G} \le 0.07$). The four \texttt{Acceleration9} fits add the two jerk components, eight data for seven parameters, and can therefore show a residual. Three reproduce the coefficients ($\chi^2_{\rm G}/\nu_{\rm G} = 0.03$, $0.58$ and $0.35$ for 04407$-$2812, 16252$-$3048 and 22374$-$4550) and 19169$-$4348 does not ($1.64$, with $\chi^2_\rho/N_{\rm rel} = 9.0$ in the resolved separations). In all, eleven of the thirteen fits that test the DR3 coefficients reproduce them. Figures~\ref{fig:set16077} and \ref{fig:set16206} show the \orvara{} diagnostics for one system of each solution type, and Appendix~\ref{app:fits} the rest of their output.

Because the largest residual in Table~\ref{tab:chisq} belongs to an \texttt{Acceleration9} fit, and the jerk enters $\chi^2_{\rm G}$ only for such solutions, we checked that the model is not propagating orbits into coefficients incorrectly. The most obvious error would be a factor of six between the DR3 jerk and the third derivative of the sky path, from the factorial convention of the Taylor expansion. There is none: the jerk recomputed analytically from the maximum-likelihood elements of the eight \texttt{Acceleration9} fits lies within a factor of $0.26$--$2.0$ of the cataloged value, and the same check on the acceleration recovers the cataloged amplitude to within $0.59$--$2.12$ for all \nfit fits and the cataloged direction to within $28\degr$ for every Mode~B fit. Exact agreement is not expected, because this check evaluates the instantaneous acceleration at J2016.0 whereas the cataloged coefficient is a least-squares fit across the DR3 window, and the two coincide only when the acceleration is nearly constant across it. Accordingly, the amplitude ratios farthest from unity, 11585$-$7754 ($2.12$), 14382$+$1402 ($1.86$) and 16077$-$2125 ($0.59$), and the one direction offset above $28\degr$, 16252$-$3048 ($50\degr$, $e = 0.87$), all belong to orbits that pass periastron inside the window, where the DR3 arc in Figure~\ref{fig:orbits} covers the most curved part of the orbit. The quantities that enter the likelihood are the finite differences across the window (Section~\ref{ssec:orvara_ext}), and these the fits reproduce: at the maximum-likelihood step the window-model acceleration of every Mode~B fit lies within $0.87$--$1.04$ of the cataloged amplitude and $5\degr$ of its direction, and the window-model jerk within $1.01$--$1.03$ for the three \texttt{Acceleration9} fits other than 14382$+$1402. The residuals in Table~\ref{tab:chisq} are therefore not attributable to an error in the orbit-to-coefficient propagation.

The two failures differ in what can be learned from them. The Mode~A fit to 19169$-$4348 has a period of $3.3$~yr, barely longer than the DR3 baseline, where a truncated sky-path model is not expected to hold (Section~\ref{ssec:gaia}), and it requires a companion mass $4.7\sigma$ above the prior that the observed $\Delta m = 0.8$~mag excludes. With three resolved epochs it cannot separate the two, and we defer it to \gaia{} DR4. The Mode~B failure, 14382$+$1402, has fifteen resolved epochs over 1.06 periods and can be taken apart, which we do in Section~\ref{ssec:14382}.

\subsection{The Failed Jerk of 14382$+$1402} \label{ssec:14382}

\begin{figure*}
\centering
\includegraphics[width=0.325\textwidth]{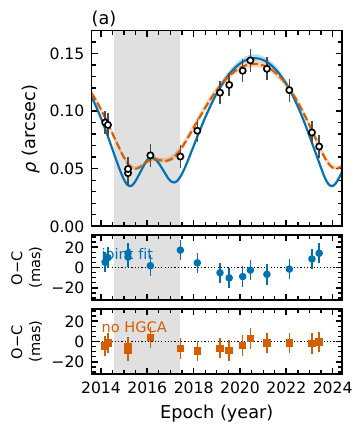}\hfill
\includegraphics[width=0.325\textwidth]{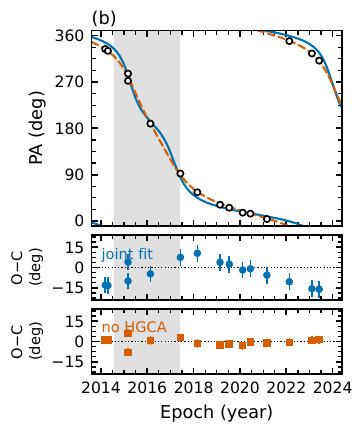}\hfill
\includegraphics[width=0.325\textwidth]{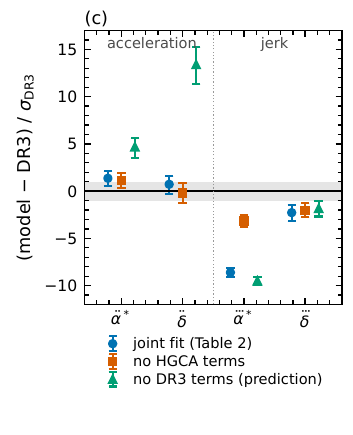}
\caption{The fit to 14382$+$1402 (HIP 71572), the system of Section~\ref{ssec:14382}. \textit{(a)}~Separation and \textit{(b)}~position angle of the resolved pair against epoch, with the fifteen WDS measures (open circles), the maximum-likelihood orbit and the 16--84\% range of the posterior draws for the Mode~B fit of Table~\ref{tab:orbits} (blue circles, solid curve) and for the diagnostic fit to the DR3 coefficients and the WDS astrometry without the HGCA terms (vermilion squares, dashed curve). The shaded band is the DR3 window, $2016.0 \pm 517.5$~d, which contains the periastron passage. The rows below give the residuals of each fit at the measured epochs, with the uncertainties used in that fit. \textit{(c)}~The DR3 acceleration and jerk coefficients: the window-model value of each fit (Section~\ref{ssec:orvara_ext}) minus the cataloged value, in units of the cataloged uncertainty as used in the fit, shown as the posterior median with its 16--84\% range. The green triangles represent predictions from fits to the HGCA and WDS astrometry without DR3 acceleration and jerk terms. Some tension remains between the HGCA-based orbit and the DR3 jerk under our treatment of the NSS coefficients. This can be further investigated with the release of Gaia DR4.
}
\label{fig:14382}
\end{figure*}

14382$+$1402 is an \texttt{Acceleration9} source whose joint fit reproduces the cataloged acceleration ($0.98$ of the amplitude, $1\degr$ from its direction, $\chi^2 = 2.8$) and fails on the jerk ($0.58$ of the amplitude, $29\degr$ off, $\chi^2 = 79$ of the $98.6$, Figure~\ref{fig:14382}c), while leaving systematic residuals of up to $16\degr$ in the resolved position angles, negative in 2014--2015, positive in 2017--2018 and negative again in 2022--2023 (Figure~\ref{fig:14382}b). The photocenter conversion is not responsible, since it divides the acceleration and the jerk by the same factor. Three diagnostic fits of the same data, seeded from the maximum-likelihood step and run with the uniform first-pass uncertainties, locate the tension. With the jerk removed, the position-angle residuals remain (rms $6\degr$, still systematic). With the \Hipparcos{} and \Hipparcos--\gaia{} proper motions removed and the DR3 coefficients kept, one orbit reproduces the fifteen resolved epochs (rms $3\degr$, no systematic pattern), the acceleration ($0.99$, $1\fdg5$) and the jerk ($0.85$, $11\degr$) together, at $i = 136\degr$ instead of $116\degr$ and $M_2 = 0.91 \pm 0.05\,\Msun$. With the DR3 acceleration and jerk removed, the \Hipparcos--\gaia{} proper motions and the resolved orbit alone predict $0.76$--$0.80$ of the cataloged acceleration in its direction to $0\fdg4$, with $M_2 = 0.77 \pm 0.05\,\Msun$, the photometric value. The DR3 coefficients are therefore consistent with the resolved orbit, and the datum in tension is the \Hipparcos{} proper motion ($\chi^2_{\rm H} = 31$ for two degrees of freedom without the DR3 terms, $15.5$ with the acceleration and $7.5$ with the jerk, whereas the \Hipparcos--\gaia{} term fits in every case). For this pair $\Delta m = 1.5$~mag gives $\fpc = 0.52$, which changes by $17\%$ per $0.3$~mag (Section~\ref{ssec:orbitfit}), and the band of the flux ratio enters the \Hipparcos{} and \Hipparcos--\gaia{} terms through $H_p$ and the DR3 terms through $G$. We cannot separate this from the quality of the \Hipparcos{} solution for a $0\farcs1$ pair.

The jerk is also the coefficient most exposed to the truncation of the sky-path model. With $P = 8.71$~yr and $e = 0.42$ the orbit passes periastron at 2016.1, inside the DR3 window, which covers a third of the period, and the acceleration of the primary changes by a factor of $2.5$--$3$ across it. Over such a window the path is not a cubic and its coefficients depend on the truncation: a cubic fitted to the model path returns a jerk $0.56$ of the cataloged value and a quintic $0.99$, while the acceleration coefficient moves from $1.1$ to $1.6$ between the cubic and the quartic. The cataloged jerk is a property of the DR3 fit to its transits rather than of the orbit, and with the largest acceleration among the Mode~B systems ($38$~mas~yr$^{-2}$ on the primary, at cataloged signal-to-noise ratios of $34$ and $116$ in its two components and $29$ and $15$ in the jerk) any such inadequacy is a large $\chi^2$. The fit is qualitatively right and statistically wrong. We therefore treat the system as a limit of the truncated sky-path model (Table~\ref{tab:orbits}, note~d) rather than as evidence against the nine-parameter solutions, tabulate the joint fit, and return to it as the test case for DR4 (Section~\ref{ssec:disc_dr4}).

\subsection{Dynamical and Photometric Companion Masses} \label{ssec:masses}

The dynamical companion masses of the Mode~B fits can be compared with the photometric estimates of Section~\ref{ssec:m2} that set $\amax$ and hence the selection of the plausible sample. Table~\ref{tab:orbits} lists the ratio $M_{2,\rm dyn}/M_{2,\rm phot}$ for the eight systems with a photometric mass, which ranges from $0.94$ to $2.03$ with a median of $1.4$. For three of the eight, 16077$-$2125, 05231$-$3142 and 11585$-$7754, the fitted $M_2$ exceeds $M_1$ ($q = 1.2$--$1.7$) although the companion is the fainter component by $0.7$--$1.0$~mag, which a main-sequence pair cannot produce. These are the three pairs with $\Delta m \le 1$~mag, and we return to them below. Two results follow.

The first concerns the dispersion. Both $\aratio$ and $M_{2,\rm dyn}/M_{2,\rm phot}$ compare the \gaia{} acceleration with a prediction from the WDS photometry, but $\aratio$ relies on the cataloged separation $\rho_1$ and the photocenter factor, whereas the orbit fit replaces both with the fitted orbit and the primary. Across these eight systems the standard deviation of $\log_{10}\aratio$ corresponds to a factor of $3.9$, whereas that of $\log_{10}(M_{2,\rm dyn}/M_{2,\rm phot})$ corresponds to a factor of $1.3$. Most of the spread in $\aratio$ therefore comes from the separation and the photocenter factor, not from the \gaia{} accelerations, which supports the use of $\aratio$ as a selection statistic but not as a per-system measurement.

The second concerns the mass scale. Without the conversion, the fit takes the photocenter's orbit for the primary's, and the companion mass it returns is too low by the factor $\fpc$ to within $0.08$ for every pair with $\Delta m \le 1.5$~mag. With the conversion, three of the five pairs with $\Delta m \ge 1.5$~mag, 03309$-$6200, 14382$+$1402 and 16268$-$4936, agree with their photometric masses to within $22\%$, comparable to the $\Delta m$ systematic of Section~\ref{sssec:m2_sys}, and 16206$+$4535 lies $39\%$ above. The factor of two for 23464$-$2302 is the break in Equation~\ref{eq:gamma}: its dynamical mass of $0.52\,\Msun$ lies on the $\gamma = 4$ branch, where the same $\Delta m$ gives $0.43\,\Msun$, whereas the photometric solution fell on the $\gamma = 2.3$ branch. The companion may also be an unresolved pair, which at fixed $\Delta m$ carries $2^{1-1/\gamma}$, or $1.5$--$1.7$, times the mass of a single star, as for Gl~229~B, whose dynamical mass exceeded that of a single object of its luminosity until it was resolved into two brown dwarfs \citep{Brandt2021_GJ229, Xuan2024}, and the two effects are not separable with the WDS photometry.

For the three pairs with $\Delta m \le 1$~mag the conversion dominates. With $\fpc = 0.25$--$0.36$ it scales the absolute astrometry by a factor of three to four, and refitting with $\fpc$ recomputed at $\Delta m + 0.3$~mag lowers $M_2$ by $26$--$41\%$. Their companion masses are bounded by the WDS magnitude difference of a $0\farcs1$ speckle pair rather than measured, and the values in Table~\ref{tab:orbits} carry that caveat. Two of the three, 05231$-$3142 and 11585$-$7754, carry the WDS \texttt{R} and \texttt{K} flags, so their cataloged $\Delta m$ is a red or infrared difference and smaller than the bolometric difference the conversion requires, which biases $\fpc$ in the observed direction (Section~\ref{sssec:m2_sys}). The sign of the offsets and their dependence on $\Delta m$ point at the photometric mass scale, not at the accelerations.

The ninth system, 08358$+$0637, has no WDS magnitude difference, so its $\fpc = 0.74$ and the companion mass in Table~\ref{tab:plausible} rest on the assumed mass ratio of Section~\ref{sssec:m2_src}. It is also the one system with an independent dynamical mass: the companion is the L5 dwarf HD~72946~B \citep{Maire2020}, with $M_2 = 69.5 \pm 0.5\,M_{\rm Jup}$ from VLTI/GRAVITY astrometry and radial velocities \citep{Balmer2023}, for which $\fpc = 1$. Refitting with $\fpc = 1$ gives $M_2 = 52_{-6}^{+9}\,M_{\rm Jup}$ with the orbit unchanged, $25\%$ below the published value, and the fit reproduces the DR3 acceleration to 1\% in amplitude and $1\degr$ in direction with about three quarters of the posterior at the published elements ($P = 16$~yr, $e = 0.50$, $i = 63\degr$). The published orbit predicts a declination acceleration of $1.63$~mas~yr$^{-2}$ for the quadratic sky-path model over the DR3 window against the cataloged $1.34 \pm 0.04$~mas~yr$^{-2}$, with the direction agreeing to $2\degr$. The acceleration of the primary changes by a factor of three across the window (periastron at 2012.2, $e = 0.5$), as for 14382$+$1402; here, however, the quadratic coefficient of the path is insensitive to the truncation ($1.59$--$1.64$~mas~yr$^{-2}$ for polynomial fits of order two to six), so the deficit is not a property of the model.

\subsection{Acceleration--Companion Alignment} \label{ssec:direction}

Of the \nplaus plausible systems, \ntest have a usable position angle. The remaining \nuntest, 08423$+$1322 and 20335$-$1333, have only occultation timings, which yield no position angle, and are untestable. Of the \ntest, \nalign ($\palign{}\%$) have the acceleration within $45\degr$ of the primary--companion axis, against the $50\%$ expected for random orientations, and the median offset is $\medoff$, against $45\degr$. A random sample of \ntest would produce an aligned fraction this high with probability $p = \binp$, and the whole distribution of offsets, weighted toward small values, differs from the uniform distribution at $D = \ksD$ ($p = \ksp$).

\begin{figure*}
\centering
\includegraphics[width=\textwidth]{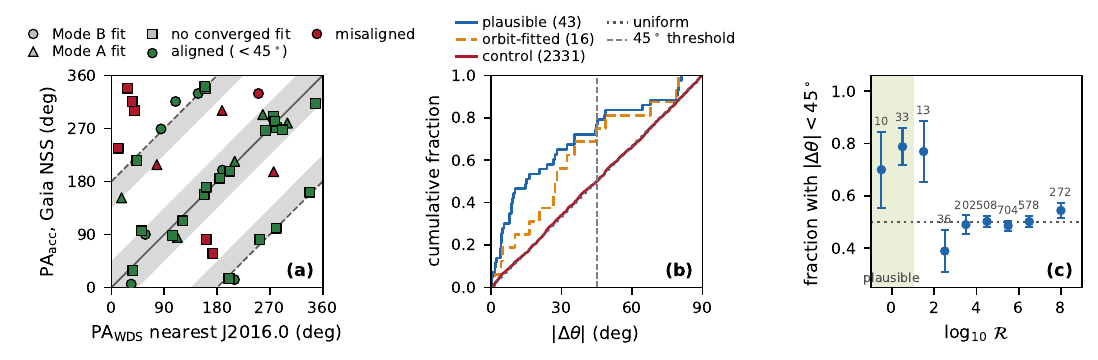}
\caption{Acceleration--companion direction test. \textit{(a)}: Position angle of the \gaia{} NSS acceleration against the WDS companion position angle at the measure nearest J2016.0, for the \ntest plausible systems with a usable position angle. The solid diagonal, $\mathrm{PA_{acc}} = \mathrm{PA_{WDS}}$, is the physical prediction of an acceleration toward the companion. The dashed diagonals are offset by $\pm180\degr$ and hold the systems whose cataloged quadrant is reversed (Section~\ref{ssec:dirtest}). The shaded bands are $\pm45\degr$ about them. \textit{(b)}: Cumulative distribution of the folded offset $|\Delta\theta|$ for the \ntest plausible systems, for the \nfittest of them with a converged fit, and for the \nctrl control systems with $\aratio > 10$. The dotted diagonal is the uniform null, and the vertical dashed line the $45\degr$ threshold. \textit{(c)}: Fraction of systems with $|\Delta\theta| < 45\degr$ in bins of a factor of ten in $\aratio$ over the \nbinned matched systems with a finite ratio, a usable position angle and $0.1 \le \aratio < 10^9$ (\nunbinned lie outside that range), with binomial standard errors and the number per bin. The shaded band is the plausible window.}
\label{fig:pa_check}
\end{figure*}

The fold hides which end of the axis the acceleration points to. Of the \nalign aligned systems, 21 lie on the solid diagonal of Figure~\ref{fig:pa_check}(a), the physical prediction, and 12 on its $180\degr$ reflection. The orbit fits resolve the ambiguity for the fitted systems, where the companion's side is fixed by the full set of measures rather than by the one nearest J2016.0. For all \Nfit fitted systems, the \nfittest in the test and the boundary case 16206$+$4535 outside it, the fitted orbit places the companion within $51\degr$ of the acceleration direction at J2016.0 (median $9\degr$, 16 of \nfit within $45\degr$), and none on the far side. Of the \nfitalign aligned fitted systems in the test, seven lie on the solid diagonal and five on the reflection, and each of the five is accounted for by the single measure the test adopts. For 11585$-$7754 and 23464$-$2302 it is the quadrant-reversed measure of Section~\ref{ssec:orbitfit}. For 16077$-$2125 it is a marginal detection at $0\farcs026$ with a $7\degr$ uncertainty that is not used in the fit, whereas the fitted orbit lies $23\degr$ from the acceleration. For 05231$-$3142 and 22374$-$4550 the adopted measure is correct but taken years after J2016.0. Each fitted orbit matches the adopted measure at its own epoch and still places the companion on the acceleration side at J2016.0, because the companion moved through half a turn in between: over 6~yr on the $P = 9.8$~yr, $e = 0.59$ orbit of 05231$-$3142, and through periastron over 3.5~yr on the $e = 0.91$ orbit of 22374$-$4550. The seven un-fitted systems on the reflection cannot be resolved individually. The direction test establishes for the population that the accelerations lie along the pair axis, and the fits establish for the \Nfit fitted systems that they point toward the companion.

A control sample confirms the null. Of the \nctrlR cross-matched systems with $\aratio > 10$, for which the cataloged companion cannot produce the acceleration, the \nctrl with a position angle (from the cataloged $\mathrm{PA}_2$, since measure files were retrieved only for the plausible sample) give \nctrlalign aligned ($\pctrl{}\%$), a median offset of $\medoffctrl$, and $D = \ksDctrl$ ($p = \kspctrl$). Alignment appears where the cataloged companion can produce the acceleration and is absent from an otherwise identical sample fifty times larger.

Figure~\ref{fig:pa_check} shows the comparison and extends it across the catalog. In panel~(a) the aligned systems cluster along the solid diagonal and its reflection in the proportion given above, and the fitted systems (circles and triangles) and un-fitted ones (squares) are distributed alike. In panel~(b) the control systems follow the uniform diagonal, whereas the plausible systems rise above it at small offsets, the departure measured by $D = \ksD$ ($p = \ksp$). Panel~(c) bins the aligned fraction by $\aratio$ in steps of a factor of ten. Alignment fills the plausible window (7 of 10 for $0.1 \le \aratio < 1$, 26 of 33 for $1 \le \aratio < 10$) and persists, less significantly, in the bin above it (10 of 13 for $10 \le \aratio < 100$, $p = 0.05$), which holds the boundary case 16206$+$4535 and near-equal pairs such as 16288$+$0559 and 08405$+$2019 that the photocenter correction moved past the window edge, so that the upper edge is conservative rather than sharp. From $\aratio = 100$ to $10^9$ the aligned fraction is consistent with $50\%$ in every bin, across seven orders of magnitude and more than two thousand systems. %A systematic in the acceleration solutions or in our analysis would not know where that boundary lies.

\begin{deluxetable*}{lrrccc}
\tabletypesize{\footnotesize}
\tablecaption{Acceleration--companion direction test by sample.\label{tab:direction}}
\tablehead{
\colhead{Sample} & \colhead{$n$} & \colhead{aligned} & \colhead{fraction} &
\colhead{median $|\Delta\theta|$} & \colhead{$p$} \\
\colhead{} & \colhead{} & \colhead{} & \colhead{} &
\colhead{(deg)} & \colhead{}
}
\startdata
Plausible, testable          & \ntest & \nalign & 0.77 & 15.0 & $\binp$ \\
\quad orbit-fitted           & \nfittest & \nfitalign & 0.75 & 27.6 & $0.038$ \\
\quad no converged fit       & \nnofittest & \nnofitalign & 0.78 &  8.8 & $3.0\times10^{-3}$ \\
\quad too sparse to fit      & \nsparsetest & \nsparsealign & 0.81 &  9.2 & $3.6\times10^{-3}$ \\
\hline
Control, $\aratio > 10$      & \nctrl & \nctrlalign & 0.50 & 45.0 & $0.50$ \\
\enddata
\tablecomments{``Aligned'' means $|\Delta\theta| < 45\degr$. $p$ is the one-sided binomial probability of the aligned fraction against the $50\%$ null. ``Too sparse to fit'' is a subset of ``no converged fit''.}
\end{deluxetable*}

Table~\ref{tab:direction} collects the statistics by subsample. The alignment is as strong among the systems without a converged fit as among the fitted ones, and the \nsparsetest systems too sparse to fit are aligned as often as the rest. For these the direction test is the only per-system check available, and it supports the cataloged companion as the perturber, whereas their acceleration ratios (median $3.06$, with 19 of the \nsparse above unity) exceed the bound because the cataloged separation of a $0\farcs1$ pair is an upper limit (Section~\ref{ssec:amax}), which raises $\aratio$ but leaves the direction untouched. The ten misaligned systems do not implicate the acceleration either. Four rest on a single measure (18060$-$4445, 22526$+$5643, 06119$-$3529 and 11181$+$1533), two are the systems set aside from fitting in Section~\ref{ssec:orbitfit_results} (20467$+$1846 and 02136$-$0849), and four are fitted systems whose adopted measure lies $0.15$--$0.30$ of an orbital period from J2016.0 (16268$-$4936, 16252$-$3048, 04407$-$2812 and 18027$-$5714), for which the fitted orbits reproduce the DR3 acceleration (Table~\ref{tab:chisq}) and place the companion within $51\degr$ of its direction at J2016.0 (Figure~\ref{fig:orbits}).

\subsection{Extending \orvara{} to \gaia{} NSS Accelerations}
\label{ssec:orvara_ext}\label{ssec:orvara_update}

The orbit fits in this work required two changes to \orvara{} \citep{orvara} and its companion package \texttt{htof} \citep{htof}: an extension of \orvara{} to use DR3 NSS acceleration solutions as a direct constraint, and an update of both packages to run on current Python and NumPy, since the published versions fail under NumPy~2 (\orvara{}) and Python~3.12 (\texttt{htof}) before any fit can complete. The updated packages are available at \url{https://github.com/t-brandt/orvara} (release v1.2.1) and \url{https://github.com/gmbrandt/HTOF} (release 1.2.1).

The absolute-astrometry input table now carries the DR3 acceleration coefficients, their uncertainties and correlation, and, for nine-parameter sources, the jerk terms, with a flag for the solution type. In Mode~A (Section~\ref{ssec:orbitfit}) the \Hipparcos{} and \Hipparcos--\gaia{} proper motion terms are assigned uncertainties large enough to remove them from the likelihood, leaving the DR3 proper motion, acceleration and jerk as the only absolute-astrometric information. In Mode~B the NSS coefficients enter together with the \Hipparcos--\gaia{} proper motion anomaly \citep{HGCA}.

The DR3 coefficients are defined on a window of $T = 1035$~d centered on J2016.0: the cataloged proper motion is the mean over the window, the acceleration is the value at its center, and the jerk is constant \citep[Eq.~7.5]{GaiaDR3doc}. \orvara{} compares the \Hipparcos, \Hipparcos--\gaia{} and \gaia{} proper motions of the model with the catalog values through Gaussian terms with the catalog covariances, marginalizing over the barycentric proper motion and the parallax \citep[Eq.~32]{orvara}. Run without epoch astrometry, as here, it evaluates the model proper motions from the sky path sampled at three epochs across each mission window. We add the acceleration and, for nine-parameter sources, the jerk as further Gaussian terms with the cataloged covariances, and evaluate them from the same sky path by finite differences over the DR3 window. For a seven-parameter source the three samples at $-T/2$, $0$ and $+T/2$ give the mean proper motion and the central acceleration exactly for a quadratic, with the jerk held at zero. For a nine-parameter source the path is sampled at $-T/2$, $-T/6$, $+T/6$ and $+T/2$, which gives all three quantities exactly for a cubic. In both cases the \gaia{} position entering the \Hipparcos--\gaia{} proper motion is the time average of the path over the window, as the catalog position is. Fitting one \texttt{Acceleration9} and one \texttt{Acceleration7} DR3 source in Mode~A and re-deriving the coefficients analytically from the posterior orbit, independently of the likelihood, returns the cataloged acceleration and jerk to well within their uncertainties.

The diagnostic figures were revised with the likelihood. In the proper-motion figure the panel below the proper-motion curve shows the model acceleration against time, the time derivative of the curve above, with the maximum-likelihood orbit in black and posterior draws colored by companion mass, and the DR3 measurement drawn over the window $2016.0 \pm 517.5$~d. For a seven-parameter solution this is a constant with its $1\sigma$ band. For a nine-parameter solution it is the segment $a + j\,(t - \mathrm{J2016.0})$, the cataloged acceleration plus the cataloged jerk times the time offset, with the envelope obtained by propagating the two uncertainties, so that the jerk is read as the slope of the segment against the model curve. The proper-motion residual is taken against the mean of the model over the window, which is the quantity the NSS solution reports, rather than against the instantaneous value at J2016.0, from which it differs by $j\,(T/2)^2/6$ for a nine-parameter source. Figures~\ref{fig:set16077} and \ref{fig:set16206} show the diagnostic for one system of each solution type.

The update to current Python leaves every result unchanged. Both packages, released as \texttt{htof}~1.2.1 and \orvara{}~1.2.1, support Python~3.10--3.13 and NumPy~2, take the parallel-tempered sampler from \texttt{ptemcee-for-orbitize} \citep[the \texttt{orbitize!} project's maintained fork of \texttt{ptemcee};][]{Vousden2016,Blunt2020}, and pass their test suites, which for \orvara{} include a closure test of the acceleration and jerk forward model and end-to-end fits. An end-to-end fit of the brown-dwarf companion HD~4747~B \citep{Crepp2016,Brandt2019}, which exercises the radial-velocity, relative-astrometry and \Hipparcos{} and \gaia{} epoch-astrometry paths together with a synthetic seven-parameter DR3 acceleration row, returns the same posterior under the original and updated stacks ($66.06 \pm 1.56$ against $66.19 \pm 1.55\,M_{\rm Jup}$ for the companion mass), within the Monte Carlo scatter of a single chain.

Because DR3 releases no epoch-level astrometry, the acceleration coefficients cannot be replaced by a forward model of the individual observations and are treated as summary statistics. The \Hipparcos--\gaia{} proper motion anomaly, which \orvara{} forward-models by integration over the mission baselines, remains the more robust of the two absolute-astrometric constraints, since the NSS coefficients are a truncated expansion whose validity degrades as the period approaches the DR3 baseline. 

\section{Discussion}\label{sec:discussion}
The results of Section~\ref{sec:results} bear on two questions: whether the DR3 acceleration solutions describe real orbital motion, and how they should be used. On the first, the direction test, the orbit fits and the comparison with the independently measured mass of HD~72946~B (08358$+$0637) agree that they do, and where the agreement is imperfect the limitation lies in the WDS data rather than in \gaia{}. On the second, the seven- and nine-parameter solutions behave differently in ways that matter for anyone fitting them, and the cataloged coefficients have a precise meaning that matters for anyone fitting them. We take the two questions in turn (Sections~\ref{ssec:disc_align} and \ref{ssec:disc_use}) and close with what \gaia{} DR4 will change (Section~\ref{ssec:disc_dr4}).

\subsection{Implications of the Direction Test}\label{ssec:disc_align}

The direction test is the one check that applies to every plausible system rather than only to the seventeen with enough relative astrometry to fit. Of the \ntest systems with a usable position angle, \nalign have the acceleration within $45\degr$ of the primary--companion axis, a $3.4\sigma$ excess over the uniform expectation of 50\% ($p = \binp$), and the full distribution of offsets rejects uniformity at $p = \ksp$. The control sample establishes that this is not a property of the analysis: the \nctrl systems with $\aratio > 10$ go through the same code and return the uniform null. A spurious acceleration from color-dependent centroiding or scan-angle systematics would not point toward a companion cataloged decades earlier by a different observer. Variability-induced motion of the photocenter \citep{Wielen1996} would, since it lies along the pair axis, but it is bounded by the brightness change it requires: $0.3$--$7$~mag over the DR3 window, more than $1$~mag for 26 of the \nplaus plausible systems and larger than any brightness change can produce for two. The seven primaries that carry the DR3 \texttt{phot\_variable\_flag} would need $0.8$--$3.5$~mag against observed $G$-band ranges of $0.02$--$0.05$~mag. Figure~\ref{fig:pa_check}(c) shows the alignment filling the plausible window $0.1 \le \aratio \le 10$, reaching the boundary case 16206$+$4535 in the bin just above it, and absent at $\aratio \ge 100$. The DR3 accelerations therefore carry directional information, including for the systems for which no orbit can be fit.

Two points follow for the population. First, \nhigher of the \nfinite matched systems with a finite ratio (\phigher{}\%) have $\aratio > 10$. Since $\amax$ is a strict upper bound, the perturber in these systems is a body other than the cataloged companion, and at a median projected separation of 2,400~AU for the cataloged pairs it must be a much closer one. An acceleration solution on a star with a known wide companion is almost never caused by that companion. Second, the spread in $\aratio$ within the plausible sample is set by the cataloged separation, not by the accelerations. Replacing $\rho_1$ with the separation implied by a fitted orbit, and the photocenter with the primary, reduces the scatter from a factor of $3.9$ to a factor of $1.3$ (Section~\ref{ssec:masses}), and the median $\aratio$ of $\medR$ sits $\medRiso$ times above the isotropic expectation of $0.65$, as it would if the cataloged $0\farcs1$ separations overestimate the true ones by a factor of two. $\aratio$ serves to select the sample, not to measure any one system.

The consequence for users of the catalog is that DR3 acceleration solutions can be taken as physical constraints on orbits, in amplitude and in direction, for both solution types. The validation here is independent of radial velocities and of \Hipparcos, so it extends to the $\sim$$3\times10^{5}$ acceleration solutions on stars with no long-baseline astrometry, where the NSS coefficients are the only absolute-astrometric information and where, as the Mode~A fits show, they constrain an orbit once a resolved companion fixes the geometry. HD~72946~B shows both sides of this for a single system: the joint fit returns a companion mass $25\%$ below the GRAVITY value, and the offset arises because the acceleration changes across the DR3 window. The limiting quantities throughout are the ones the WDS supplies, the separation and the magnitude difference, rather than the ones \gaia{} supplies.

\subsection{The Use of DR3 Acceleration Solutions} \label{ssec:disc_use}

Three caveats apply to the DR3 acceleration solutions. First, the coefficients are the derivatives of the sky path at J2016.0, not orbit-averaged quantities, and they lose their meaning when the acceleration changes appreciably across the 34-month baseline, whether because the period is short or because an eccentric orbit passes periastron inside the window. The \texttt{significance} field helps here in one respect. For \texttt{Acceleration9} it is the signal-to-noise ratio of the jerk (Section~\ref{ssec:gaia}), so a high value flags exactly such a system, as it does for 02136$-$0849, the system in our sample whose acceleration is least stable across the baseline, and for 14382$+$1402 (\texttt{significance} 44, in the top fifth of \texttt{Acceleration9} solutions), the one fitted system whose jerk the joint fit cannot reproduce. It cannot be used to compare the precision of the acceleration between the two solution types, for which the coefficient uncertainties themselves serve (Section~\ref{ssec:ratio}). 11585$-$7754 illustrates the limitation. Its periastron falls at 2016.75, inside the window \citep{Tok2024a}. Started from a generic initial orbit, the fit settled on a solution with $P \sim 10^{3}$~yr that matched the DR3 acceleration but not the resolved astrometry. Only when started from the published elements did it find the 14.9-yr orbit of Table~\ref{tab:orbits}, which matches both. That orbit is not the published one: with seven resolved epochs the relative astrometry admits both the 28-yr, $0\farcs113$ orbit of \citet{Tok2024a} and the 14.9-yr, $0\farcs081$ orbit of Table~\ref{tab:orbits}, and the DR3 acceleration together with the \Hipparcos--\gaia{} proper motions selects the shorter one. The acceleration alone would not have selected the right solution. 08358$+$0637 is a milder case. With periastron 2.4~yr before the window and $e = 0.5$, the acceleration changes by a factor of three across it, and the companion mass of the fit is 25\% below the published dynamical mass (Section~\ref{ssec:masses}).

Second, we find no evidence that nine-parameter accelerations are less reliable than seven-parameter ones. In the joint \Hipparcos--\gaia{} fits, where the two types can be compared on equal terms, eight of nine reproduce the DR3 coefficients, seven with $\chi^2_{\rm G}/\nu_{\rm G} \le 0.2$, for both types alike (Table~\ref{tab:chisq}). The exception, 14382$+$1402, reproduces the acceleration and not the jerk. It passes periastron inside the window, where the cubic coefficients of the path are not converged, and its jerk is reproduced once the \Hipparcos{} proper motion is removed (Section~\ref{ssec:orbitfit_results}). Of the four \gaia{}-only fits that test their coefficients, three reproduce them, and the one that does not, 19169$-$4348, has a period of $3.3$~yr, barely longer than the baseline. Both failures belong to the truncated sky-path model, not to one solution type. Fits of NSS coefficients without an independent long-baseline constraint test the model weakly, since a seven-parameter \gaia{}-only fit with two resolved epochs has fewer data than parameters.

Third, for a pair that \gaia{} does not resolve, the cataloged acceleration is that of the photocenter, which is smaller than the primary's by the factor $\fpc = |1-\beta/B|$ of Equation~\ref{eq:photocenter}. For the plausible pairs with $\Delta m \le 1.5$~mag the factor is $0.14$--$0.52$, and it is sensitive to the magnitude difference. An error of $0.3$~mag in $\Delta m$ changes $\fpc$ by $17\%$ at $\Delta m = 1.5$~mag and by $30$--$45\%$ at $\Delta m \le 1$~mag, and that change carries directly into any companion mass derived from the acceleration (Section~\ref{ssec:masses}). A dynamical mass from an NSS acceleration for a companion within $\sim$2~mag of its primary requires this conversion and a flux ratio in a known band, and is limited by the precision of that flux ratio.

Our sample carries no information about the low-acceleration regime relevant to tests of modified gravity \citep{Milgrom1983}, since the requirement that the companion produce a detectable acceleration excludes the weak-field limit by construction.

\subsection{Prospects for \gaia{} DR4} \label{ssec:disc_dr4}

The first two caveats of Section~\ref{ssec:disc_use} share one cause, which is that DR3 publishes the coefficients of a truncated sky-path model and not the epoch astrometry behind them. With epoch astrometry in DR4 the model can be refit rather than adopted, so that the truncation is tested instead of assumed, orbit-averaged quantities can replace the instantaneous ones that make $\aratio$ ill-defined for systems such as 11585$-$7754, and the seven- and nine-parameter models can be compared on the same source. Epoch data also allow such fits for sources that DR3 modeled with five parameters \citep{156_systems}, enlarging the sample, and the longer DR4 baseline extends the sensitivity toward the longer periods of resolved pairs. 14382$+$1402 is the case to start with: an $8.7$-yr orbit with periastron inside the DR3 window, a fifteen-epoch resolved orbit that predicts the sky path, and a jerk that the truncated model cannot define. The third caveat, the photocenter, DR4 does not remove, since it is set by the photometry of the pair rather than by the astrometry.

DR4 will not remove the other limit on this test, which is the resolved astrometry. The \nsparse plausible systems without an orbit lack WDS measures, not \gaia{} data, and one or two speckle epochs at $0\farcs1$--$0\farcs3$ for each would turn their direction-only validations into orbit and mass determinations. The \nsparsealign of them already aligned are the natural targets.

\section{Conclusion}\label{sec:conclusion}

We have tested the \gaia{} DR3 NSS acceleration solutions against the Washington Double Star catalog. Cross-matching \nnss acceleration solutions yields \nuniq pairs within $2\arcsec$, and for each we compare the measured acceleration with $\amax$, the face-on Newtonian bound on the acceleration the cataloged companion can impart on the \gaia{}-tracked photocenter, through the ratio $\aratio = \ameas/\amax$.

For \phigher{}\% of the sample the cataloged companion cannot be responsible, since $\aratio > 10$ against a strict upper bound. At the median projected separation of the matched pairs the companion's pull is orders of magnitude below the DR3 detection threshold, so an NSS acceleration on a star with a known wide companion is almost never caused by that companion. The window $0.1 \le \aratio \le 10$ isolates \nplaus close pairs for which it can be. Their cataloged separations are quantized at the WDS resolution limit, which makes $\aratio$ a selection statistic rather than a per-system measurement and, by the projection bound $\aratio = \sin^3\theta$, shows those separations to be overestimated by about a factor of two.

Orbit fits with \orvara{}, extended to accept the DR3 acceleration and jerk coefficients as direct constraints and run with the absolute astrometry converted from the photocenter to the primary, converge for \nfitplaus of the \nplaus systems and one boundary case. Of the 13 fits that test the DR3 coefficients, 11 reproduce them, seven- and nine-parameter solutions alike, and the two that fail, 14382$+$1402 and 19169$-$4348, are the two whose acceleration changes substantially within the DR3 baseline. Replacing the cataloged separation with the fitted orbit and the photocenter with the primary reduces the scatter between dynamical and photometric companion masses from a factor of $3.9$ to a factor of $1.3$, so the dispersion in $\aratio$ comes from the WDS inputs and not from the accelerations. The dynamical masses agree with the photometric ones where the photocenter conversion is well determined, and for HD~72946~B, the one companion with an independent dynamical mass, the fit recovers the published orbit and a mass $25\%$ below the GRAVITY value.

The direction test is independent of the fits. Of the \ntest plausible systems with a usable position angle, \palign{}\% have the acceleration within $45\degr$ of the primary--companion axis, with a median offset of $\medoff$, whereas \nctrl control systems with $\aratio > 10$ return the uniform 50\%. The test uses direction alone, so it is free of the separation and mass systematics that limit the amplitude comparison and it applies to the \nsparse systems too sparsely observed for an orbit. Within the range where the cataloged companion can produce the acceleration, the DR3 accelerations point at it.

Epoch astrometry in \gaia{} DR4 will allow the acceleration model to be refit rather than adopted and the two solution types to be compared on the same source. For this test, however, the limiting catalog is the WDS. One or two speckle epochs at $0\farcs1$--$0\farcs3$ for each of the \nsparse systems without an orbit would turn their direction-only validations into orbit and mass determinations. The updated \orvara{} and \texttt{htof} are available at \url{https://github.com/t-brandt/orvara} and \url{https://github.com/gmbrandt/HTOF}.

\section*{Acknowledgments}

This work has made use of data from the European Space Agency (ESA) mission \gaia{} (\url{https://www.cosmos.esa.int/gaia}), processed by the \gaia{} Data Processing and Analysis Consortium (DPAC; \url{https://www.cosmos.esa.int/web/gaia/dpac/consortium}). Funding for the DPAC has been provided by national institutions, in particular the institutions participating in the \gaia{} Multilateral Agreement.

This work has made use of the Washington Double Star Catalog maintained at the U.S.\ Naval Observatory. We thank the many observers who have contributed measurements to the WDS over more than two centuries, and the USNO for supplying the WDS measure files used in this work.

This publication makes use of data products from the Two Micron All Sky Survey, which is a joint project of the University of Massachusetts and the Infrared Processing and Analysis Center/California Institute of Technology, funded by the National Aeronautics and Space Administration and the National Science Foundation.

This research has made use of the VizieR catalogue access tool and the cross-match service provided by CDS, Strasbourg, France.

\facilities{Gaia \citep{GaiaMission}, Hipparcos \citep{Hipparcos_1997}, 2MASS \citep{2MASS}}

\software{\orvara{} \citep[v1.2.1, this work;][]{orvara}, \texttt{htof} \citep[v1.2.1, this work;][]{htof}, astropy, numpy, scipy, matplotlib, emcee, ptemcee-for-orbitize \citep{Vousden2016,Blunt2020}}
\appendix
\section{The two example fits}\label{app:fits}

\begin{figure*}
\centering
\includegraphics[width=0.34\textwidth]{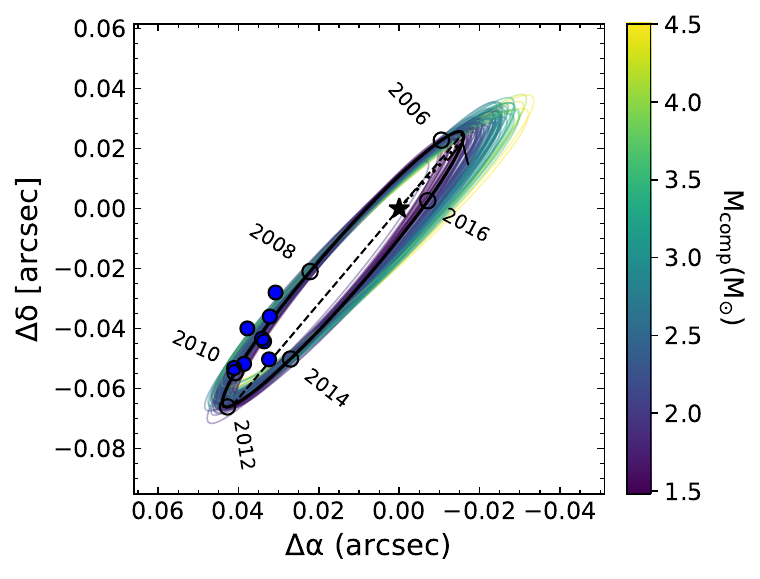}\hfill
\includegraphics[width=0.31\textwidth]{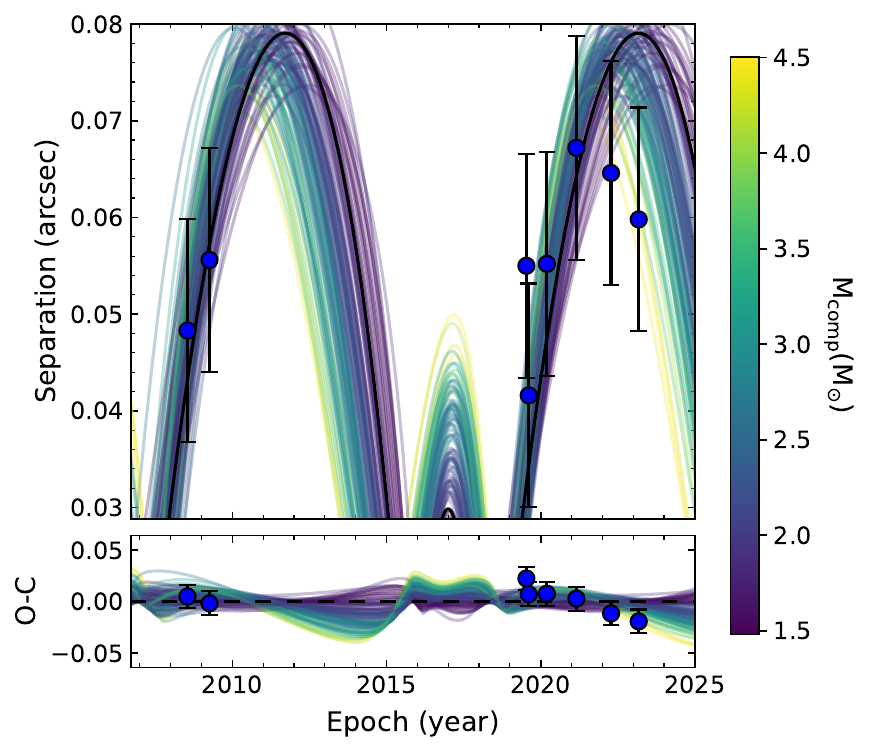}\hfill
\includegraphics[width=0.31\textwidth]{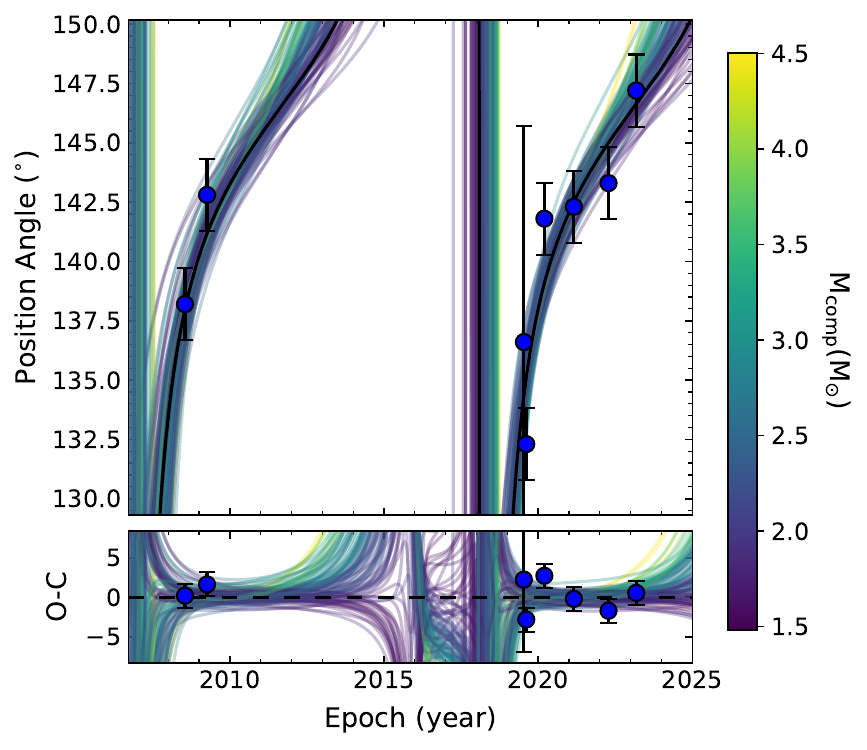}\\[6pt]
\includegraphics[width=0.85\textwidth]{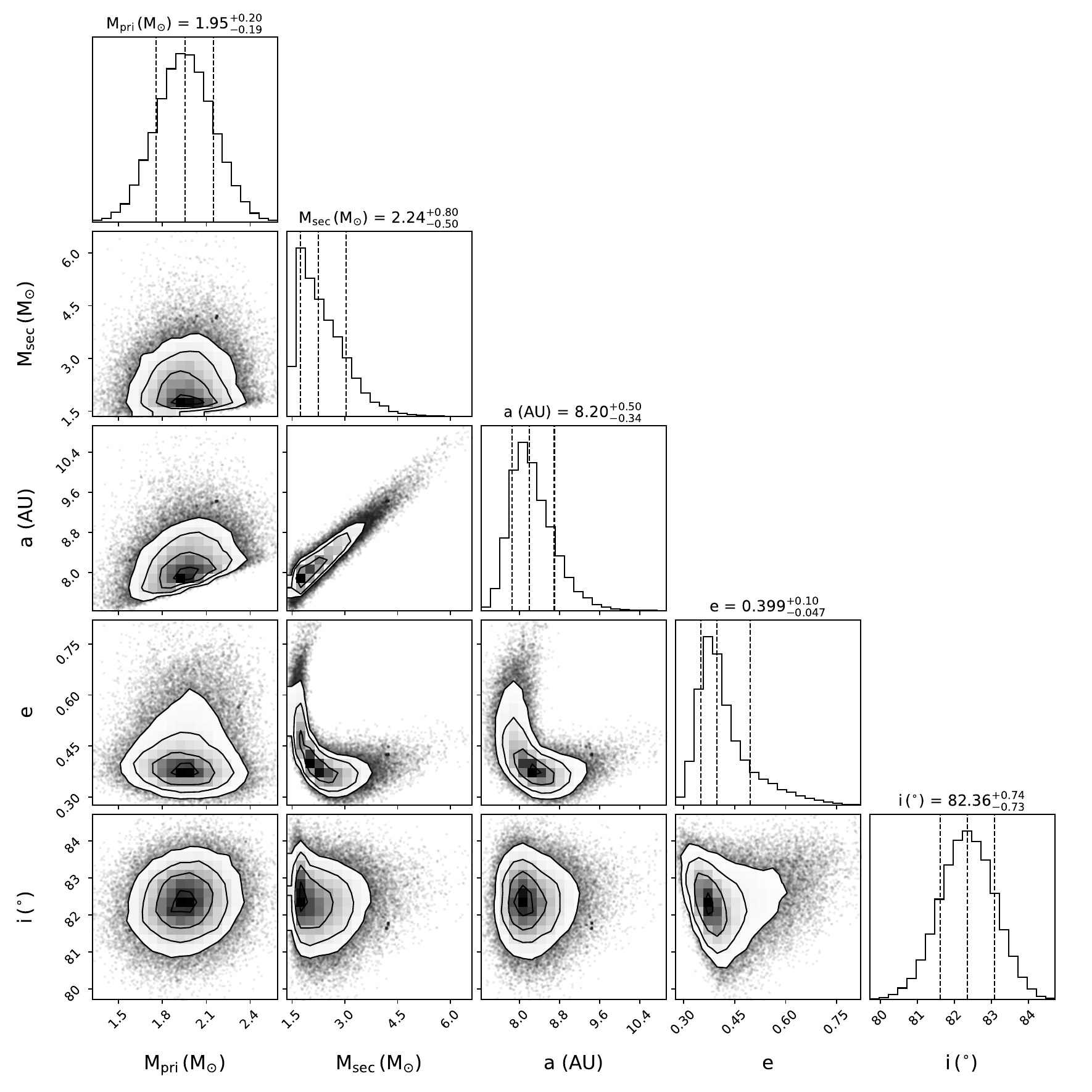}
\caption{The fit of 16077$-$2125 (\texttt{Acceleration7}). \textit{Top:}~relative orbit, separation and position angle with their residuals. \textit{Bottom:}~posterior. The companion mass is degenerate with the semimajor axis and the eccentricity along the ridge that the short arc leaves open.}
\label{fig:app16077}
\end{figure*}

\begin{figure*}
\centering
\includegraphics[width=0.34\textwidth]{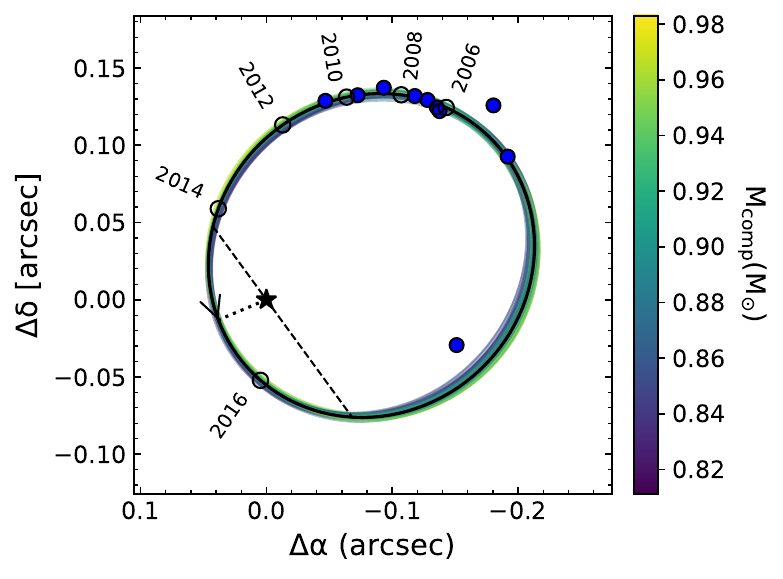}\hfill
\includegraphics[width=0.31\textwidth]{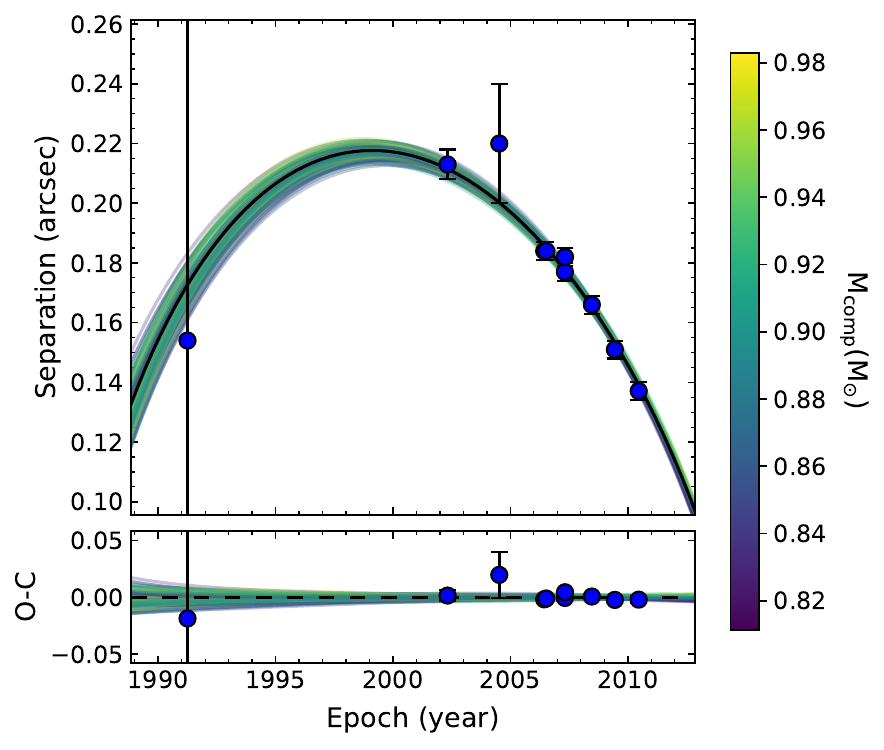}\hfill
\includegraphics[width=0.31\textwidth]{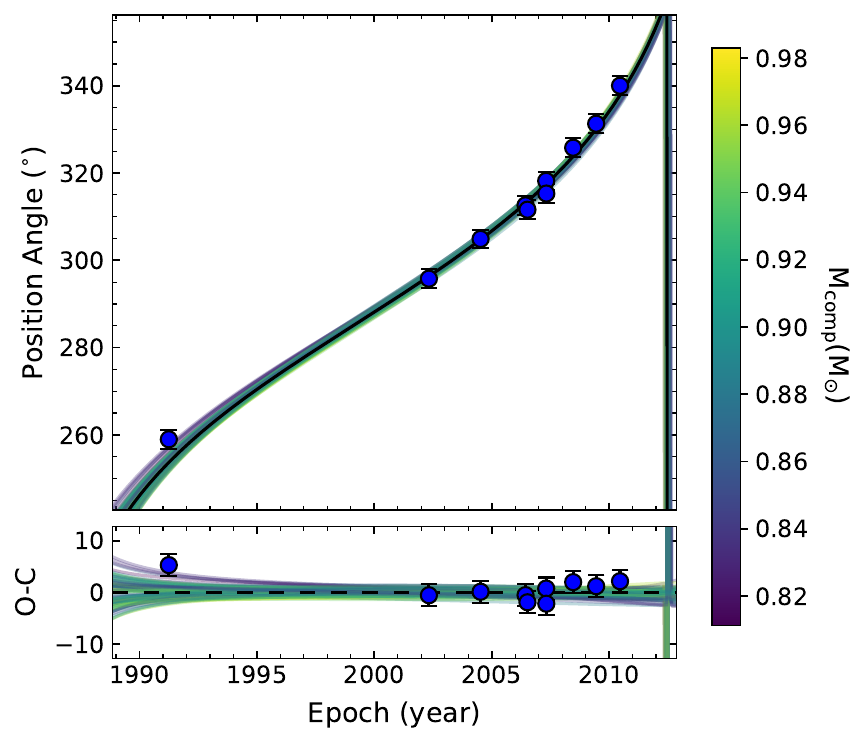}\\[6pt]
\includegraphics[width=0.85\textwidth]{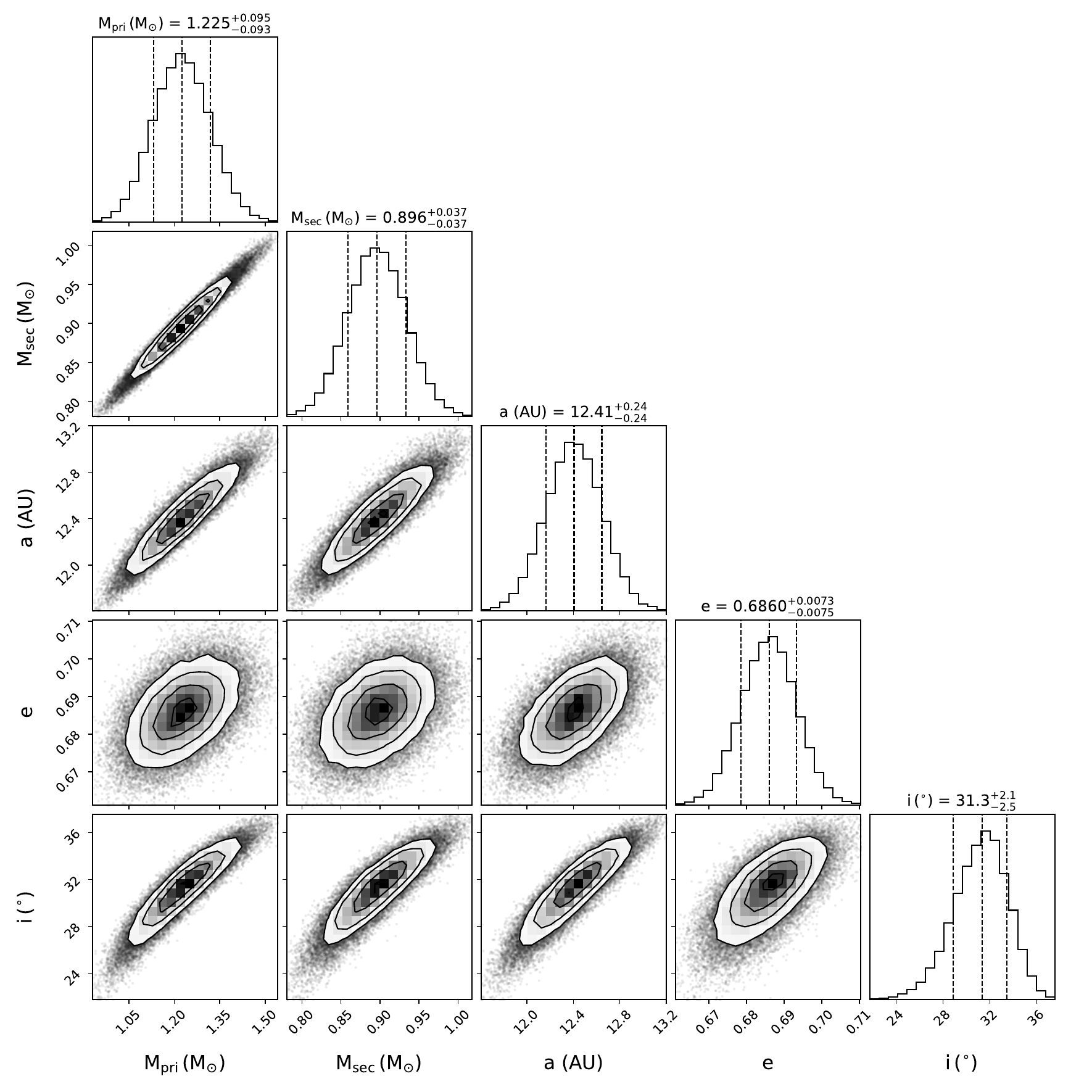}
\caption{As Figure~\ref{fig:app16077}, for 16206$+$4535 (\texttt{Acceleration9}).}
\label{fig:app16206}
\end{figure*}

Figures~\ref{fig:app16077} and \ref{fig:app16206} complete the \orvara{} output for the two fits of Figures~\ref{fig:set16077} and \ref{fig:set16206}: the relative orbit with the WDS measures and the epochs marked, the separation and position angle against epoch with their residuals, and the corner plot of the posterior for the primary and companion masses, semimajor axis, eccentricity and inclination, with the 16th, 50th and 84th percentiles marked. The maximum-likelihood orbit is black and the posterior draws are colored by companion mass.

\clearpage

\clearpage

\bibliography{references}
 
\end{document}